\documentclass[12pt,a4paper]{iopart}

\usepackage{ifpdf}
 \ifpdf
   	\usepackage[T1]{fontenc}  
   	\usepackage[pdftex]{hyperref,graphicx}	
    	\hypersetup{colorlinks,%
    	citecolor=blue,%
    	filecolor=green,%
    	linkcolor=blue,%
    	urlcolor=blue,%
   	}
  	\usepackage{pdfpages} 
 \else
 	\usepackage[T1]{fontenc} 
 	\usepackage[dvips]{graphicx}		
 	\usepackage[active]{srcltx} 
	\newcommand{\url}[1]{$ #1 $} 
	\newcommand{\href}[2]{$ #2 (#1) $} 
 \fi

\newcommand{\half}[0]{\frac{1}{2}}
\newcommand{\bigO}[0]{\mathcal{O}}
\newcommand{\SSzero}{\rm SS$_0$}
\newcommand{\SSpi}{\rm SS$_\pi$}
\newcommand{\Rzero	}{$R_0$}
\newcommand{\Rpi}{$R_\pi$}
\newcommand{\Mone}{$S_1$}
\newcommand{\Mtwo}{$S_2$}

\newcommand{\rhorel}{\rho_\textrm{\tiny rel}} 

\usepackage{iopams}

\usepackage[utf8x]{inputenc}
\usepackage[normalem]{ulem}
\usepackage{amssymb}
\usepackage{amsopn}

\newcommand{\eqref}[1]{\eref{#1}}

\newcommand{\text}{\textrm}

\DeclareMathOperator{\Imp}{Im}

\renewcommand{\S}{\mathbf{S}}

\renewcommand{\Im}{\Imp}

\usepackage[english]{babel}

\usepackage{xcolor}
\usepackage{cite}
\begin{document}

\title[Basins of Attraction for Chimera States]{Basins of Attraction for Chimera States}

\author{Erik A.~Martens${}^{a,b}$, Mark J.~Panaggio${}^{c,d}$, Daniel M.~Abrams${}^{d,e,f}$}
\address{
${}^a$Dept. of Biomedical Sciences, University of Copenhagen, Blegdamsvej 3, 2200 Copenhagen, Denmark}
\address{
${}^b$Dept. of Mathematical Sciences, University of Copenhagen, Universitetsparken 5, 2200 Copenhagen, Denmark
}
\address{
${}^c$Mathematics Dept., Rose-Hulman Institute of Technology, Terre Haute, IN}
\address{
${}^d$Dept. of Engineering Sciences and Applied Mathematics, Northwestern University, Evanston, IL}
\address{
${}^e$Dept. of Physics and Astronomy, Northwestern University, Evanston, IL}
\address{
${}^f$Northwestern Institute on Complex Systems, Northwestern University, Evanston, IL
}
\ead{erik.martens@ds.mpg.de}

\vspace{10pt}
\begin{indented}
\item[]\today
\end{indented}

\begin{abstract} 
Chimera states---curious symmetry-broken states in systems of identical coupled oscillators---typically occur only for certain initial conditions. Here we analyze their basins of attraction in a simple system comprised of two populations. Using perturbative analysis and numerical simulation we evaluate asymptotic states and associated destination maps, and demonstrate that basins form a complex twisting structure in phase space.  Understanding the basins' precise nature may help in the development of control methods to switch between chimera patterns, with possible technological and neural system applications.
\end{abstract}


%
\vspace{2pc}

%
\submitto{\NJP}
%
\maketitle
%
%

\section{Introduction}

Self-emergent synchronization is a key process in networks of coupled oscillators, and is observed in a remarkable range of systems, including pendulum clocks, pedestrians on a bridge locking their gait, Josephson junctions, flashing fireflies, the beating of the heart, circadian clocks in the brain, chemical oscillations, metabolic oscillations in yeast, life cycles of phytoplankton, and genetic oscillators~\cite{Huygens1967ab,Strogatz2005,Wiesenfeld1998,Elowitz2000,Buck1968, Peskin1975,Michaels1987,Liu1997,Kiss2002,Massie2010, Ghosh1971,Dano1999,Taylor2009}. 
About a decade ago, a study~\cite{Kuramoto2002} revealed the existence of chimera states, in which a population of identical coupled oscillators splits up into two parts, one synchronous and the other incoherent.  This state is counter-intuitive as it appears despite the oscillators being identical. Recent experiments using metronomes, (electro-)chemical oscillators and lasing systems~\cite{MartensThutupalli2013,Wickramasinghe2013,Tinsley2012,Hagerstrom2012,Schonleber2014} have demonstrated the existence of chimera states in real-world settings; previous theoretical studies have also confirmed the robustness of chimeras subjected to a range of adverse conditions, including additive noise, varied oscillator frequencies, varied coupling topologies, and other imperfections~\cite{Shima2004,Omelchenko2008,Panaggio2013,Maistrenko2015,Abrams2004,Feng2015,Laing2009,Laing2012a, Panaggio2015_2, Panaggio2015_1,Pikovsky2008}.

\begin{figure}[htp]
  \centering 
  \includegraphics[width=0.6\textwidth]{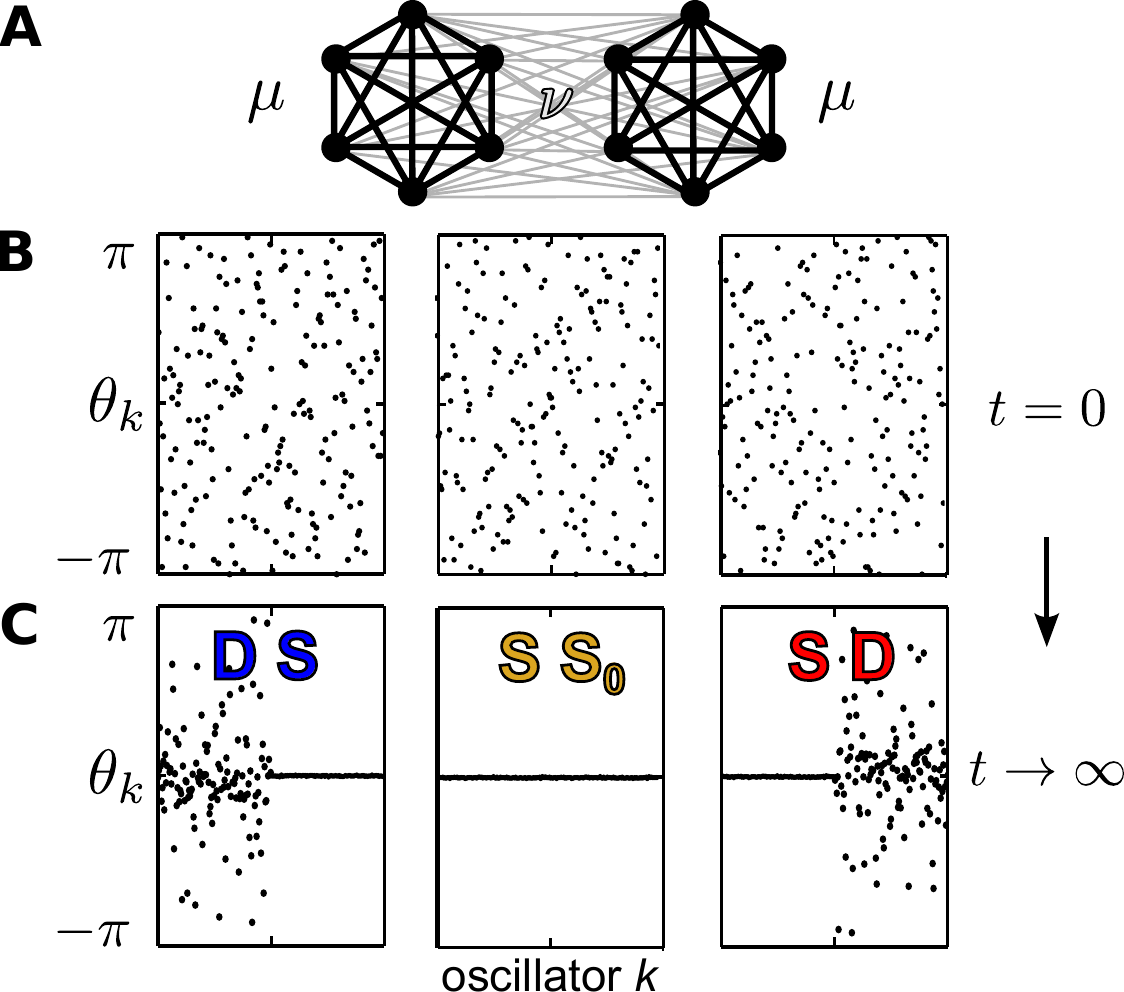}
  \caption{\label{figure1}
  {\bf A}: Schematic of the system under investigation. {\bf B}: Three superficially similar oscillator phase distributions taken as initial conditions.  {\bf C}:  Oscillator phase distributions after long-time evolution of system---each corresponds to the initial condition shown directly above it. DS: ``desync-sync'' state; \SSzero: ``sync-sync'' state; ``SD'': ``sync-desync'' state.
  }
\end{figure}

Chimeras are known to arise in systems with nonlocal coupling that decays with increasing distance between phase oscillators, thus bridging the gap between the extremes of local (nearest-neighbor) and global (all-to-all) coupling\footnote{For oscillators with non-constant amplitudes it appears that local and global coupling can both be sufficient~\cite{Laing2015,SethiaSen2014,Schmidt2014}.}. Such long-range coupling is characteristic of many real-world technological and biological~\cite{Phillips1993,Swindale1980,JdMurray2002} systems. 
In many systems, chimeras are steady-state solutions stably coexisting with the fully synchronized state\footnote{Known exceptions where chimera states and fully synchronized states are not necessarily bi-stable are certain generalizations of the present system with amplitude dynamics~\cite{Sethia2013,Schmidt2014} or non-linear delay feedback~\cite{Omelchenko2008,Yeldesbay2014}.}, not emerging via spontaneous symmetry breaking, and are thus only attained via a certain class of initial conditions~\cite{Kuramoto2002,Abrams2004,Panaggio2015_1}. 
Figure~\ref{figure1} graphically demonstrates this puzzling aspect of basins of attraction for chimera states: apparently similar initial conditions (panel B) can evolve to completely different steady-states (panel C).
Thus, a natural question arising in any practical situation is: given a random initial phase configuration, how likely is the system to converge to a chimera state? Even though this important question was raised in 2010~\cite{Motter2010}, basins of attraction for chimera states have not yet been investigated systematically.

One difficulty with the examination of basins of attractions is that they are computationally expensive to obtain, e.g.~via Monte Carlo simulation~\cite{Menck2013}. Here, in contrast, we use primarily analytic methods to explain the structure of the phase space and provide a systematic study of the basins of attraction leading to chimeras in the thermodynamic limit.

\textit{Model.} The simplest realization of nonlocal coupling is achieved with two populations, where each population is more strongly coupled to itself than to the neighboring population (see Figure~\ref{figure1} panel A). It has been used as a model for several investigations of chimera states ~\cite{Abrams2008,Montbrio2004,Laing2009,Laing2012,Laing2012a,MartensThutupalli2013,OlmiMartens2015}; here chimeras manifest themselves as a state with one synchronous and one asynchronous population.  Accordingly, we consider the Kuramoto-Sakaguchi model with $n=2$ populations~\cite{Abrams2008,Montbrio2004} each of size $N^{\sigma}$, 
\begin{eqnarray}\label{eq:goveqns1}
 \dot{\theta_k^{\sigma}} &=& 
 \omega + \sum_{\sigma'=1}^2 \frac{K_{\sigma\sigma'}}{N^{\sigma'}}\sum_{l=1}^{N^{\sigma'}}
 \sin{(\theta_l^{\sigma'}-\theta_k^\sigma-\alpha)},\
\end{eqnarray}
where $\theta_k^{\sigma}$ is the phase of the $k$th oscillator $k= 1, \ldots, N^{\sigma}$ in population $\sigma \in \{1,2\}$ and $\omega$ is the oscillator frequency. For consistency with previous work~\cite{Abrams2008, Montbrio2004, Martens2010var}, we assume the coupling is symmetric with neighbor-coupling $K_{\sigma\sigma'}=K_{\sigma'\sigma}=\nu$ and self-coupling $K_{\sigma\sigma}=\mu$. Imposing without loss of generality $\mu+\nu=1$, the coupling can be parameterized by the coupling disparity $A=\mu-\nu$. We redefine the phase lag parameter via $\beta=\pi/2-\alpha$ as chimeras emerge in the limit of near-cosine-coupling  ($\beta\rightarrow0$) for this type of system~\cite{Abrams2008, Martens2010bistable, Martens2010var}. The mean field order parameter $R_{\sigma}e^{i\Phi_{\sigma}} = (N^\sigma)^{-1}\sum_{j=1}^{N^{\sigma}}\exp{(i\theta^{\sigma}_j)}$ describes the  synchronization level of population $\sigma$  with $R_{\sigma}=1$ for perfect and $R_{\sigma}\leq 1$ for partial synchronization.
We consider the thermodynamic limit $N^{\sigma}\rightarrow\infty$, allowing us to express the ensemble dynamics in terms of the continuous oscillator density $f^{\sigma}(\theta,\omega)$. This facilitates a low-dimensional description of the dynamics via the Ott-Antonsen (OA) ansatz~\cite{Ott2008b, Ott2009, Ott2011} in terms of the mean-field order parameter of each population, 
$\rho_{\sigma}(t)e^{i\phi_{\sigma}(t)}=\int e^{i\theta}f^{\sigma}(\theta,t)d\theta$
with $0<\rho_\sigma \leq 1$, see~\ref{app:OA} and~\ref{app:goveqns2pop}.

By virtue of the translational symmetry $\phi_\sigma\rightarrow\phi_\sigma+\text{const.}$, the resulting dynamics are effectively three dimensional with the angular phase difference $\psi=\phi_1-\phi_2$, obeying 
\begin{eqnarray}\label{eq:goveqnsrhopsi1}
 \dot{\rho}_1 &=& \frac{1-\rho_1^2}{2}\left[\mu \rho_1\sin{\beta} + \nu \rho_2 \sin{(\beta-\psi)} \right]\\\label{eq:goveqnsrhopsi2}
 \dot{\rho}_2 &=& \frac{1-\rho_2^2}{2}\left[\mu \rho_2\sin{\beta}+\nu \rho_1\sin{(\beta+\psi)}  \right]\\\nonumber
 \dot{\psi} &=& \frac{1+\rho_2^2}{2\rho_2}\left[\mu \rho_2\cos{\beta} + \nu\rho_1\cos{(\beta+\psi)} \right]\\\label{eq:goveqnsrhopsi3}
 &-& \frac{1+\rho_1^2}{2\rho_1}\left[\mu\rho_1 \cos{\beta} + \nu\rho_2\cos{(\beta-\psi)} \right],\
\end{eqnarray}
on domain $D = \{(\rho_1,\rho_2,\psi)|0 < \rho_{1,2}\leq 1, -\pi\leq \psi \leq \pi\}$. 

\begin{figure}[tp!]
  \centering
  \includegraphics[width=0.6\textwidth]{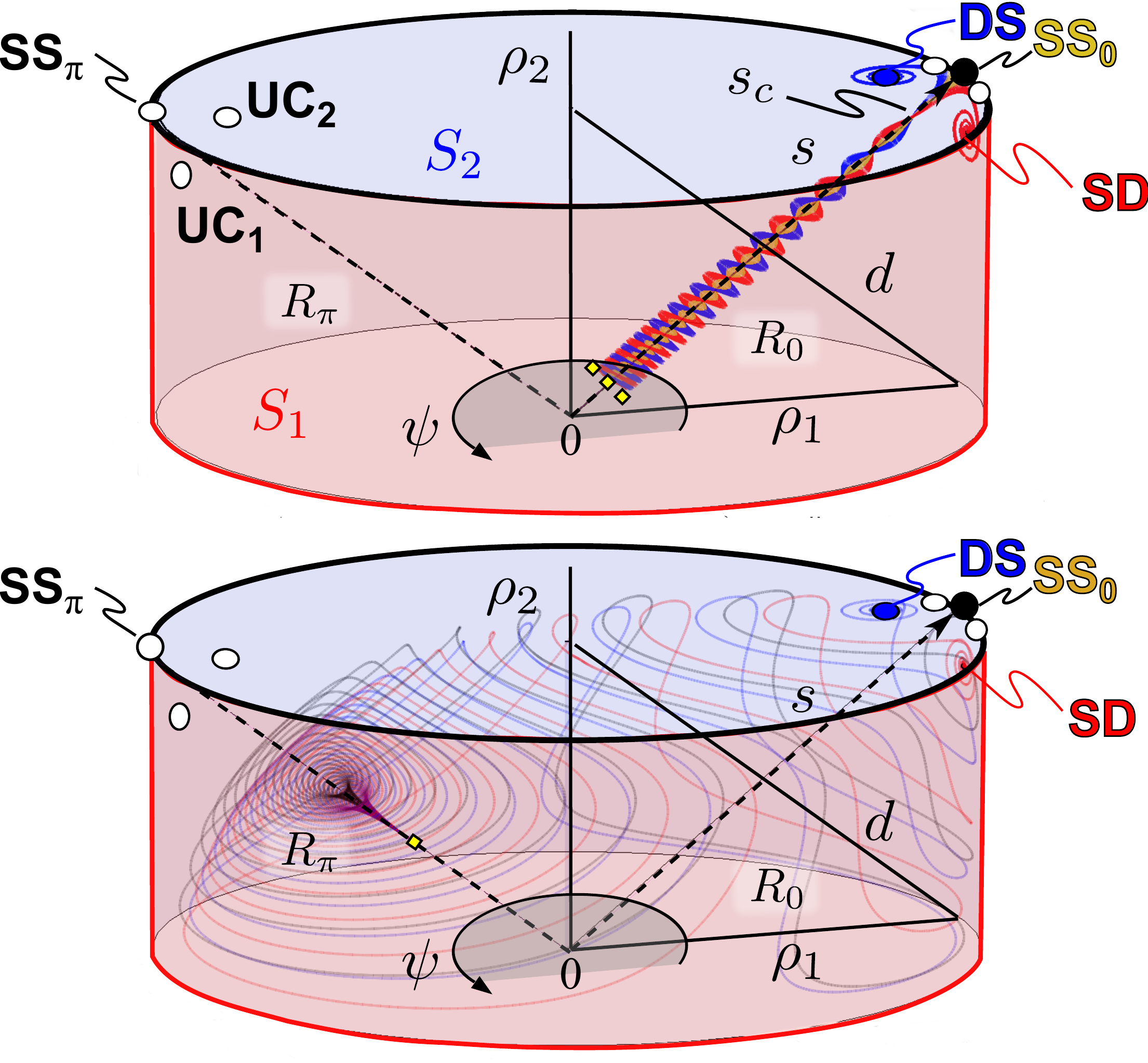}
  \caption{\label{figure2} State variables $(\rho_1,\psi,\rho_2)$ are interpreted as cylindrical coordinates.  Phase space is structured by (i) two invariant rays, $R_0$ and $R_\pi$ (dashed); and (ii) two invariant surfaces, \Mone\, and \Mtwo, forming the side and top surfaces of the cylinder slab. Except for a set of measure zero, all trajectories converge to one of three locations: SD chimera state on \Mone  \,(red), DS chimera state on \Mtwo\, (blue), or fully synchronized state \SSzero~(yellow).  $(A,\beta)=(0.1,0.025)$; filled/empty circles denote stable/unstable fixed points. Small yellow dots denote initial conditions.
}
\end{figure}

Phase space is visualized using cylindrical coordinates $(\rho_1,\psi, \rho_2)$, see Figure~\ref{figure2}.  The translation $\Pi:\beta \mapsto \beta+\pi$ reverses time in Eqs.~(\ref{eq:goveqnsrhopsi1})-(\ref{eq:goveqnsrhopsi3}), thus inverting flow in phase space and stability of fixed points; we restrict our attention to $0\leq \beta\leq \pi$ in what follows. 
\section{Invariant Manifolds and fixed points}
Analysis of equations \eqref{eq:goveqnsrhopsi1}-\eqref{eq:goveqnsrhopsi3} reveals the existence of two invariant surfaces defined by $S_\sigma=\{(\rho_1,\rho_2,\psi)|\rho_\sigma=1\}\subset D$ (the top (blue) and lateral (red) surfaces of the cylinder displayed in Figure~\ref{figure2}). The dynamics on these manifolds were studied previously~\cite{Abrams2008}: chimera states are born in a saddle-node bifurcation and undergo a Hopf bifurcation for larger coupling disparity $A=\mu-\nu$. The resulting stable limit cycle grows with $A$ until eventually it is destroyed in a homoclinic bifurcation. Studying the basins of attraction, we generalize the previous analysis by considering the entire three-dimensional phase space $D$.  

Numerically, we observe that all trajectories with initial conditions $\rho_1,\rho_2<1$ are attracted to one of the invariant surfaces. From there, any of three attractors can be asymptotically approached: (i) a partially synchronized limit point (stable chimera, either SD or DS), (ii) a limit cycle (breathing chimera, either SD or DS), or (iii) the fully synchronized state \SSzero~at $(\rho_1,\rho_2,\psi)=(1,1,0)$. Furthermore, unstable fixed points exist: a fully synchronized state \SSpi~at $(\rho_1,\rho_2,\psi)=(1,1,\pi)$, and several unstable saddle chimeras (UC) (see~\cite{Panaggio2015_2} and~\ref{app:fixedpoints}).  The dynamics on \Mone\, and \Mtwo\, are related due to the invariance of Eqs.~(\ref{eq:goveqnsrhopsi1})-(\ref{eq:goveqnsrhopsi3}) under the symmetry operation $\Sigma: (\rho_1,\rho_2,\psi)\mapsto(\rho_2,\rho_1,-\psi)$. 

Outside of \Mone\, and \Mtwo, trajectories follow a complex winding motion, structured around the two invariant rays \Rzero ~and \Rpi~defined by $\rho_1=\rho_2$ with $\psi=0$ and $\psi=\pi$, respectively (see Figure~\ref{figure2} and~\ref{app:IM}). Other than the origin, which is a repeller, there are no fixed points for $\rho_1,\rho_2<1$ (see~\ref{app:fixedpoints}). Thus, limit cycles in the interior of the phase space are also absent. In principle, a chaotic attractor could appear inside $D$ but is not observed. 

\section{Numerical investigations}\label{sec:destinationmaps}
\begin{figure}[htp!]
\label{SIfigure3}
 \centering
 \includegraphics[width=0.7\textwidth]{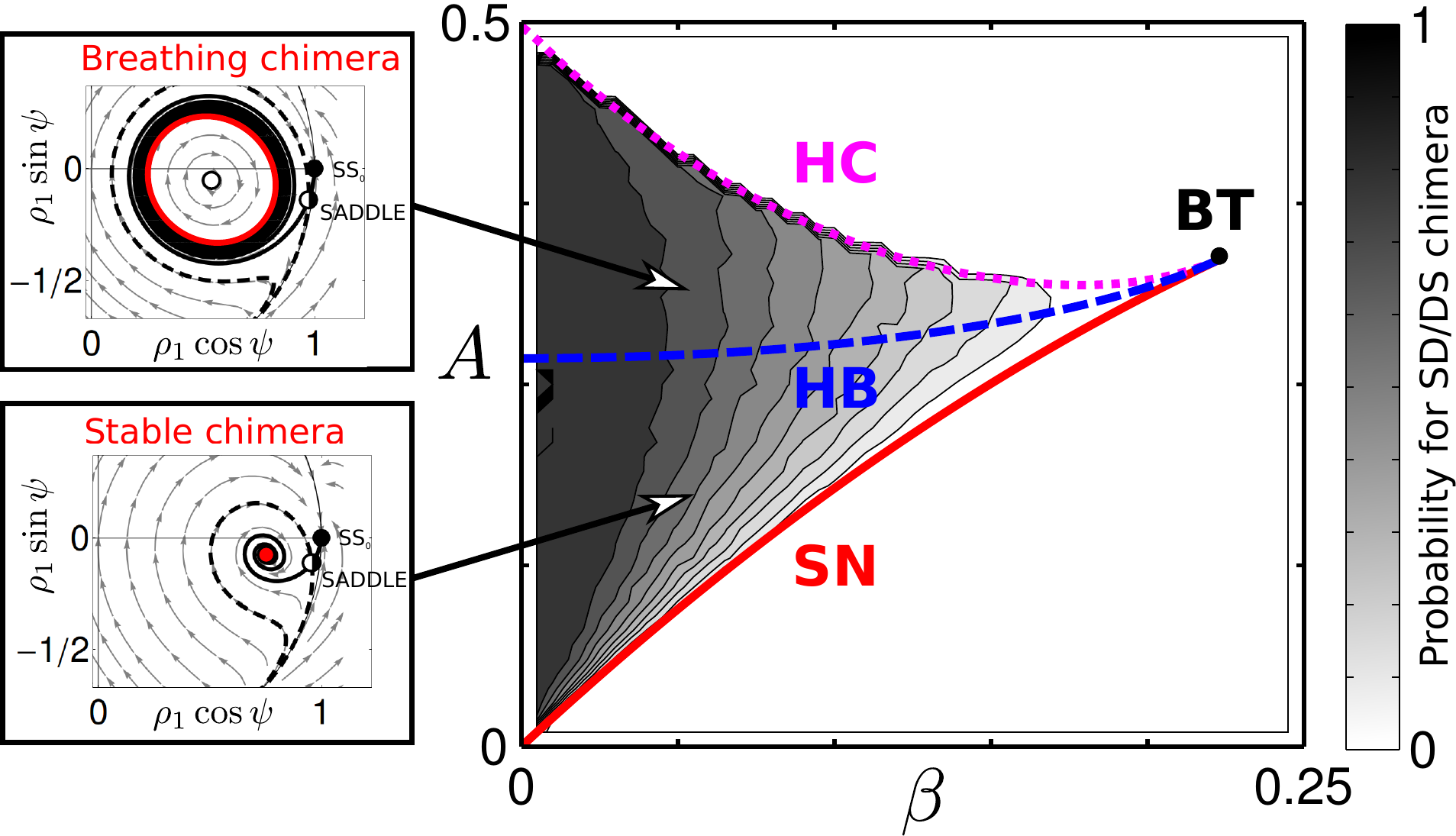}
 \caption{  
 Probabilities to obtain chimera states via random sampling of initial conditions $(\rho_1,\rho_2,\psi)$. Chimeras appear within the wedge defined by a saddle-node bifurcation (SN, solid) for small $A$ and a homoclinic bifurcation (HC, dotted) for large $A$~\cite{Abrams2008}.
 Phase portraits in the $\rho_2=1$ plane  are shown (insets) with stable nodes (full circles), unstable chimera (white circle) and saddle chimera (half-filled circles), together with its stable (solid) and unstable (dashed) manifolds. 
 For intermediate $A$, the asynchronous order parameter undergoes a Hopf bifurcation (HB, dashed). Probabilities for ending up in either SD/DS chimera were\ measured by realizing 1000 random initial conditions $(\rho_1,\rho_2,\psi)$ for each parameter value set.
 Further details are in~\ref{app:numerics}.
 }
\end{figure}
First insights regarding basins of attraction for chimera states were gathered via simple Monte Carlo integration of uniformly distributed random initial conditions for $\rho_1,\rho_2\in[0,1]$ and $\psi\in[-\pi,\pi]$. These computations reveal that the probability $p(A,\beta)$ of ending up in a chimera state depends primarily on $\beta$ with a maximum value for $\beta\rightarrow 0$, see Figure~\ref{SIfigure3}. This approach provides information about the sizes of the basins of attraction, but it reveals little about their structure. We therefore ask: how is the three-dimensional phase space structured?

To better reflect symmetries of the phase space, the dynamics may be re-expressed in terms of the sum and difference of the order parameters (see Figure~\ref{figure2}), 
$s = \frac{1}{2}(\rho_1 + \rho_2)$ with $0\leq s\leq 1$, and $d = \frac{1}{2}(\rho_1 - \rho_2)$,
with $-a(s)\leq d \leq a(s)$ where $a(s)=\frac{1}{2}-|\frac{1}{2}-s|$~(see Eqns.~\eqref{eq:sdpsi_eqnA}-\eqref{eq:sdpsi_eqnC}). 


\begin{figure}[htp!]
  \centering
  \includegraphics[width=0.6\textwidth]{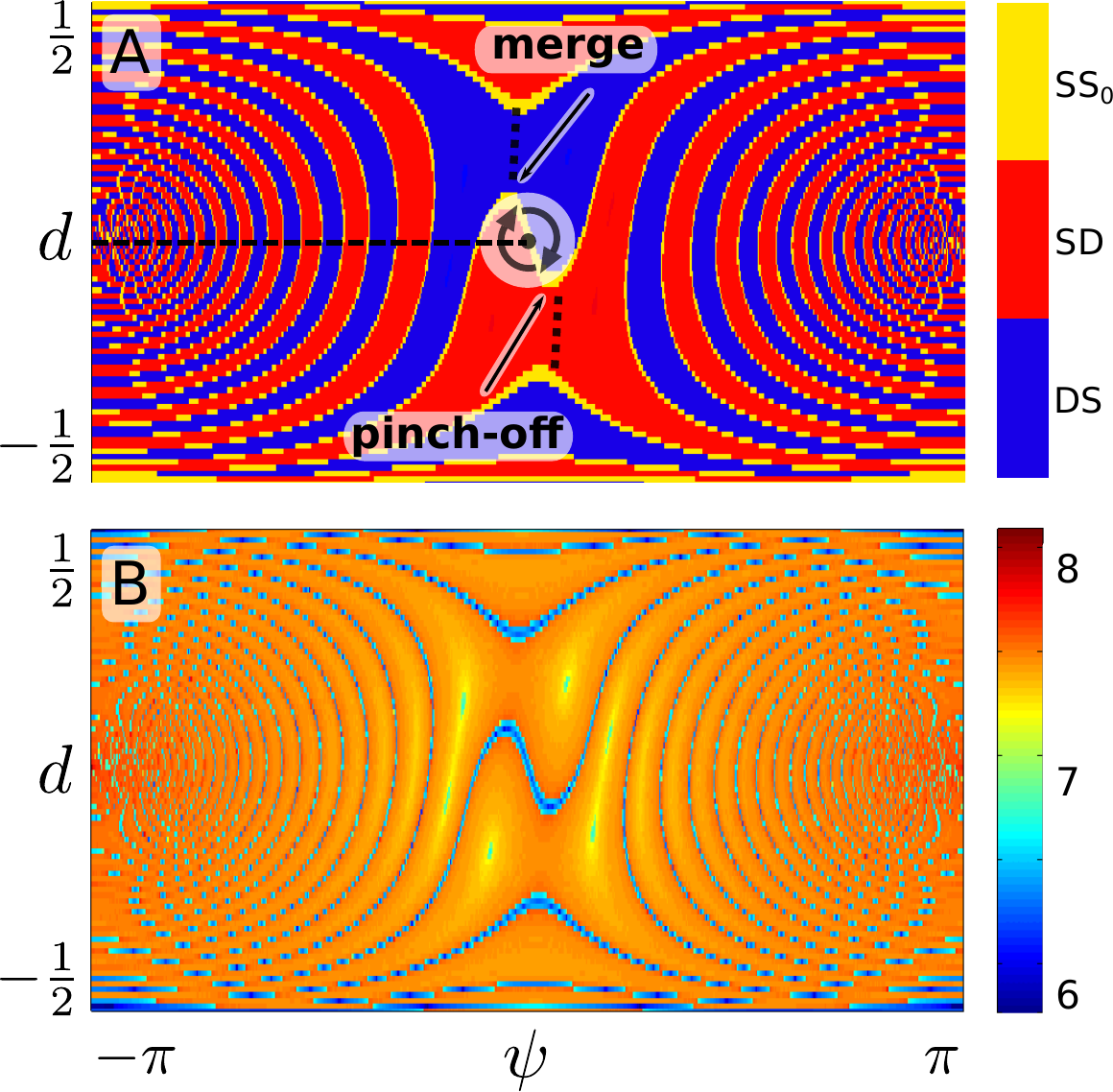}
  \caption{
  \label{figure3}
  ({\bf A}) Destination map section in the $(d,\psi)$-plane with $s=0.56625<s_c$, for SD (red), DS (blue) and \SSzero~(yellow) states.   When $s$ increases, the basin boundaries perform a spiraling motion as indicated by arrows. ({\bf B}) The logarithmic times $\log{T}$ to destination reflect the structure of the destination map in ({\bf A}). Times peak at the interface boundaries between \SSzero~and SD/DS regions (see also Figure~\ref{figure4}).  Parameters are $(A,\beta)=(0.1,0.025)$.
  }
\end{figure}
 
In order to characterize the structure of the basins of attraction, we compute the destination maps for a set of initial conditions $(s,d,\psi)$.  Figure~\ref{figure3}A shows a typical cross-section of the destination map with fixed $s$: basins form a spiraling structure around $R_\pi$ (the ray $(d,\psi)=(0,\pi)$), with SD and DS basins always separated by the (often thin) basin for \SSzero. The thickness of the basin spiral arms increases away from $R_\pi$, with maximum near $R_0$ (the ray $(d,\psi)=(0,0)$). 

The area ratio between basins for SD (or, by symmetry, DS) and \SSzero~is related to the probability that a random initial condition will lead to a chimera state, and depends on parameters $A$ and $\beta$ as follows. For $\beta \to 0$, the \SSzero~basin occupies an infinitesimal fraction of the area.  As $\beta$ increases, the \SSzero~basin increases its area until it occupies the entire plane at $\beta=\beta_{\rm SN}(A)$ when the chimera state is annihilated through a saddle-node bifurcation. For $A < A_{\rm SN}(\beta)$ or $A > A_{\rm HC}(\beta)$, no chimera state exists and the entire basin belongs to the \SSzero~state.  With increasing $A>A_{\rm SN}$, the (total) basin area of \SSzero~gradually decreases from 100\% approaching a constant near the homoclinic bifurcation, see Figure~\ref{SIfigure1}.

As $s$ increases from zero, basins merge and pinch-off in an alternating fashion (see Figure~\ref{figure3}, Sec.~\ref{sec:analysis} and Supplementary Video 1) so that the basin boundaries rotate clockwise about $R_0$ ($(d,\psi)=(0,0)$ in Figure \ref{figure3}A).  Once $s$ reaches $s_c \approx \sqrt{1-A}$, this rotation stops, demonstrating that knowledge of the trajectory position in the $s=s_c$ plane is sufficient for determining its final fate.

\begin{figure}[htp!]
  \centering
  \includegraphics[width=0.6\textwidth]{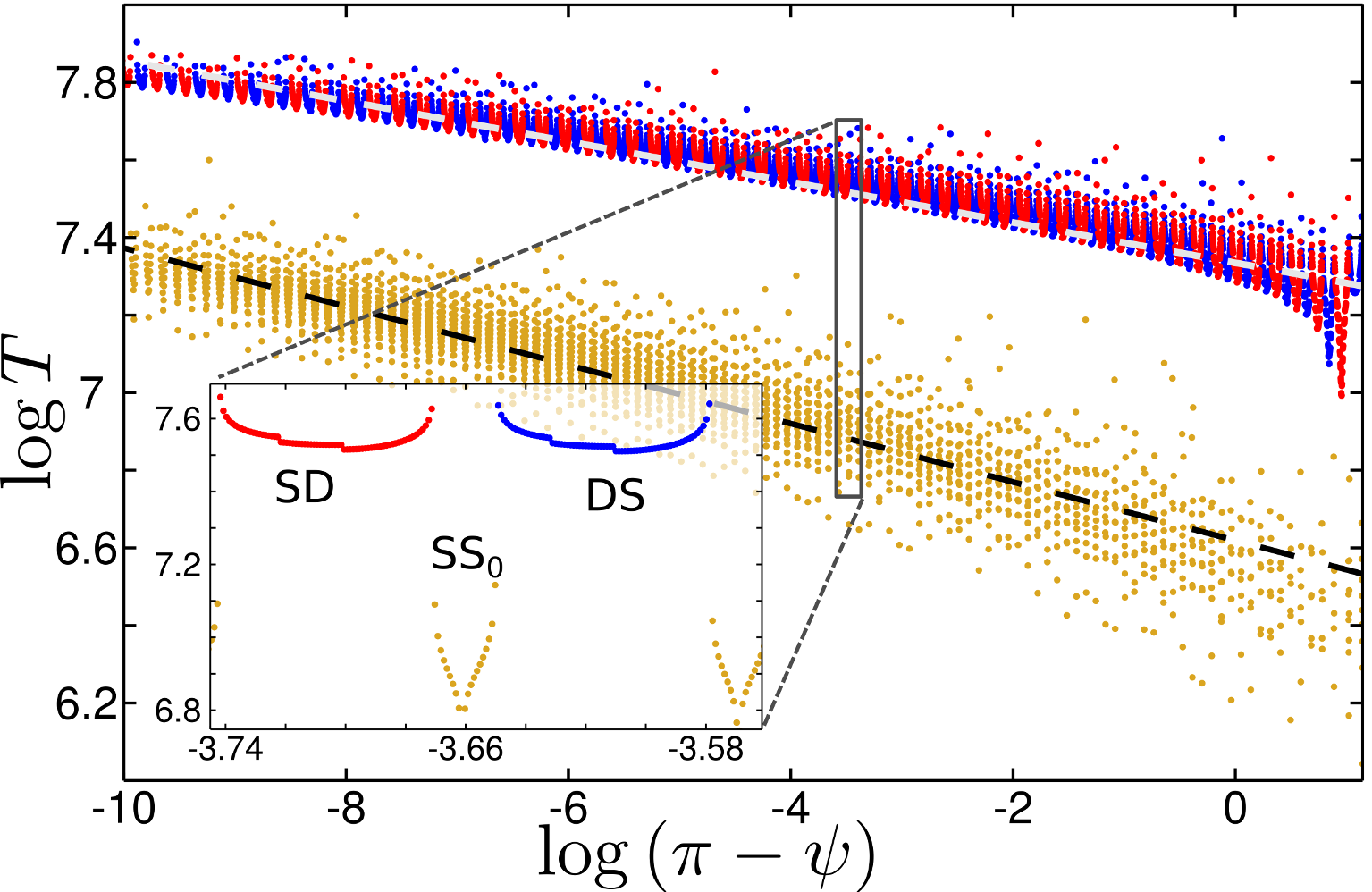}
  \caption{\label{figure4}
  Times $T$ to reach $\epsilon$-neighborhood of fixed points \SSzero/DS/SD; trajectories start from $(s,d,\psi) = (0.56625,0,\psi)$ for $-\pi\leq\psi\leq\pi$, $A=0.1,\beta=0.025$ (straight line on Figure~\ref{figure3}). Average destination times $T$ grow like a power law as $\psi \rightarrow \pm \pi$, thus basin structure is self-similar around $(d,\psi)=(0,\pi)$. $T$ diverges at the boundary between SD/DS and \SSzero~basins (inset), since these trajectories lie on stable manifolds leading to saddle points on invariant manifolds $S_1, S_2$. 
  }
\end{figure}

The basin density appears singular near $R_\pi$, with a nested structure that allows even tiny perturbations of the initial condition to strongly influence the final state.  More generally, the highly alternating basin structure is reflected in the times to reach steady-state attractors, which are displayed in Figure~\ref{figure3} B. Figures~\ref{figure4} and~\ref{SIfigure2} show destination times $T$ along a section of that figure between the origin and $(d,\psi)=(0,\pi)$, revealing a power-law behavior that may result from this nested spiral arm basin structure.

Figure~\ref{figure4} also reveals that destination times diverge on the basin boundaries (see inset), which is explained by the fact that these boundaries form separatrix sheets: these are the two-dimensional stable manifolds emanating from the saddle chimeras on \Mone\, and \Mtwo, originating in the saddle-node bifurcation that gives birth to chimeras, see Figure~\ref{figure2} and~\cite{Abrams2008}. Numerical continuation of those sheets (see Figs.~\ref{SI_Separatrices1} and~\ref{SI_Separatrices2}, and~\ref{app:numerics_cont}) displays the same twisting motion as seen in Figure~\ref{figure3}.

\section{Analysis}\label{sec:analysis}
A complete analysis of the basins for the entire phase space is difficult to achieve, but the main features of the basin structure have their origin in the invariant rays $R_0$ and $R_\pi$ about which a perturbation analysis can be made for small $A$ and $\beta$ (asymptotic results remain qualitatively in fair agreement for parameter values further off the origin).

\subsection{Perturbation analysis around the invariant ray \texorpdfstring{\Rzero}{R0}}\label{sec:R0perturbation}
We consider the coupling constants $(\mu, \nu)$ to be perturbed from global coupling ($A=0$) by setting $\mu = \half (1+A)$, $\nu = \half (1-A)$ as in \cite{Abrams2008}.  We then make the perturbative approximation that $\psi$, $d$, $\beta$ and $A$ are all small and of the same order (while keeping in mind that $A > 2 \beta$ is required for the existence of a chimera state in this limit). This means that near $R_0$, we make the ansatz~\cite{Abrams2008}:  
\begin{eqnarray*}
  \psi &= \psi_1 \epsilon + \bigO(\epsilon^2), \\
  d &= d_1 \epsilon + \bigO(\epsilon^2), \\
  \beta &= \beta_1 \epsilon + \bigO(\epsilon^2), \\
  A &= A_1 \epsilon + \bigO(\epsilon^2)~.
\end{eqnarray*}
After making a change of variables $x=d_1$, $y=\half s\psi_1$\footnote{The change of variables is chosen so that that the spiraling cycles become circular in shape.}, and then a second change of variables $x=r\cos(\theta)$ and $y=r\sin(\theta)$ we find the following equations:
\begin{eqnarray}
\frac{dr}{dt}&=-\left(\half A_1\sin(2\theta)+\beta_1\right)s^2r\epsilon,\label{eq:r_R0}\\
\frac{d\theta}{dt}&=\half\left(1-s^2\right)-\half\left[1+\cos(2\theta)s^2\right]A_1\epsilon,\label{eq:th_R0}\\
\frac{ds}{dt}&=\half s\left(1-s^2\right)\beta_1\epsilon.\label{eq:s_R0}
\end{eqnarray}
Note that the derivatives of $r$ and $s$ are both order $\epsilon$ while the derivative of $\theta$ is order 1 (when $s$ is not close to 1). Thus $\theta$ evolves on a fast time scale while $r$ and $s$ evolve slowly.  We may therefore use the method of averaging on the higher order terms involving $\theta$ to simplify these equations to
\begin{eqnarray}
\label{eq:rdotnearR0avg}
\frac{dr}{dt} &= -\beta_1s^2r\epsilon,\\
\frac{d\theta}{dt} &= \half\left(1-s^2-A_1\epsilon\right), \label{eq:thdotnearR0avg}\\
\frac{ds}{dt} &= \half s\left(1-s^2\right)\beta_1\epsilon \label{eq:sdotnearR0avg}.\
\end{eqnarray}

This reveals a couple of properties.  First of all, as expected, solutions will be spirals around the \Rzero~manifold. The radius $r$ goes to 0 as $t\rightarrow\infty$ (for the truncated equations).  Thus, trajectories slowly converge toward the~\Rzero~manifold until the approximations break down when $\frac{d\theta}{dt}=0$ and higher order terms become significant. In other words, the \Rzero-manifold is weakly attracting.   
The frequency of rotation is $\omega=\half(1-s^2-A_1\epsilon)$.  
So, as $s \rightarrow s_c \approx \sqrt{1-A}$, the rotation frequency $d\theta/dt\rightarrow 0$. This is here referred to as the $s_c$-plane. In this plane, the trajectories cease to have a spiral character and instead begin to separate and evolve toward the fully synchronized state or the DS or SD chimeras. 

Note that there is an alternative way that the ``critical plane''  could be defined.  Setting $d\theta/dt=0$ in Eq.~\eqref{eq:th_R0} yields a minimum $s$ solution $s_c=1-A_1\epsilon + \bigO(\epsilon^2)$ (possible only for particular $\theta$ values), thus $s_c \approx 1-A$ is the smallest value of $s$ for which rotation about ray $R_0$ may stop.
The difference between the two expressions $s_c \approx \sqrt{1-A} \approx 1-A/2$ and $s_c \approx 1-A$ comes from whether averaging has been applied or not.  In the former case (averaged equations), rotation about $R_0$ stops on average over all $\theta$; in the latter case, rotation about $R_0$ stops only for \textit{some particular $\theta$}.

By symmetry, if a trajectory originating at $(s,d,\psi)$ converges to the SD state, the trajectory originating at $(s,-d,-\psi)$ must converge to the DS state. Therefore, the position relative to a separating boundary in the $s_c$ plane determines the final state.  Numerical integration confirms that trajectories converging to SD and DS chimeras form opposing sides of a positively oriented double helix centered on \Rzero~(red and blue in Figure~\ref{figure2} and Supplementary Video 2). 

The ``rotation'' of the basin boundary as $s$ increases along the $R_0$ manifold (indicated symbolically in Figure~\ref{figure3} A) can also be understood analytically in the perturbative limit close to $R_0$. 
Equation \eqref{eq:sdotnearR0avg} can be solved explicitly to get 
\begin{eqnarray}\label{eq:R0_s_evolution}
  s(t) &=& s_o \Big/ \sqrt{s_0^2 + (1-s_0^2)e^{-\epsilon \beta_1 t}}~.
\end{eqnarray}
Taking $\dot{\theta} = (1-s^2)/2$ to lowest order (from \eqref{eq:thdotnearR0avg}), we can substitute in for $s(t)$ and then integrate from $t=0$ to $t=t_{\rm crit}$ to approximate the total angle change about the $R_0$ manifold over the course of the trajectory.  Here $t_{\rm crit}$ represents the time at which the trajectory $s(t)$ reaches the critical plane $s=s_c$:
\begin{eqnarray}\label{eq:R0tcrit}
  t_{\rm crit} &=& \frac{1}{\epsilon \beta_1} \ln\left( \frac{1-s_0^2}{s_0^2} \frac{(2-A)^2}{A(4-A)}  \right) = \frac{1}{\epsilon \beta_1} \ln \left( \frac{(1-s_0^2) s_c^2}{(1-s_c^2)s_0^2} \right).
\end{eqnarray}
Integrating $\dot{\theta}$ gives a total angle change
\begin{eqnarray}\label{eq:R0_rotationangle}
  \Delta \theta &=& \frac{1}{\epsilon \beta_1} \ln \left( \frac{1-A/2}{s_0} \right) = \beta^{-1} \ln(1-A/2) - \beta^{-1} \ln s_0~.
\end{eqnarray}
(This can also be written as $\Delta \theta = \beta^{-1} \ln(s_c/s_0)$, and then this expression is valid for either definition of $s_c$.)
Thus, the boundary angle is proportional to $\beta^{-1} \ln(1-A/2) - \beta^{-1} \ln(s)$, yielding a rotation rate of $(\beta s)^{-1}$ as the section plane $s$ varies uniformly.

Since the angle of a trajectory at the critical plane determines the basin the trajectory belongs to, the appearance of the basin boundary in a section plane orthogonal to the ray $R_0$ is just a line with angle proportional to $\Delta \theta$.

\subsection{Perturbation analysis around the invariant ray \texorpdfstring{\Rpi}{Rpi}}\label{sec:Rpiperturbation}
We can perform a similar analysis around the $R_\pi$ ray  by a similar ansatz:
\begin{eqnarray*}
  \psi &=& \pi+\psi_1 \epsilon + \bigO(\epsilon^2), \\
  d &=& d_1 \epsilon + \bigO(\epsilon^2), \\
  \beta &=& \beta_1 \epsilon + \bigO(\epsilon^2), \\
  A &=& A_1 \epsilon + \bigO(\epsilon^2)~.
\end{eqnarray*}
This time we make an (analogous) change of variables $x=d_1$, $y=\half\frac{s\sqrt{1-s^4}}{1+s^2}\psi_1$\footnote{The change of variables is chosen so that that the spiraling cycles become circular in shape; the particular shape differs from the one near the $R_0$-manifold.}, and again convert to polar coordinates.  This yields 
\begin{eqnarray}
\frac{dr}{dt}&=\left[\half\beta_1\left(1-s^2\cos{(2\theta)}\right)-\half s^2\sqrt{\frac{1-s^2}{1+s^2}}\sin{(2\theta)} \right]r\epsilon\label{eq:r_Rpi}\\\nonumber
\frac{d\theta}{dt}&=-\half\sqrt{1-s^4}+\left[\half\sqrt{\frac{1-s^2}{1+s^2}}\left(1-s^2\cos{(2\theta)}\right)A_1 +\half\beta_1\sin{(2\theta)}s^2 \right]\epsilon,\label{eq:th_Rpi}\\
\frac{ds}{dt}&=\left[\half s\left(1-s^2\right)\beta_1 A_1+\frac{1+s^2}{4s}\sqrt{\frac{1+s^2}{1-s^2}}r^2\sin(2\theta) \right]\epsilon^2.\label{eq:s_Rpi}
\end{eqnarray}
Again we find that the derivative of $\theta$ is larger than the other derivatives, allowing us to reduce the system to
\begin{eqnarray}
\frac{dr}{dt}&=\half\beta_1r\epsilon,\\
\frac{d\theta}{dt}&=-\half\sqrt{1-s^4}+\left[\half\sqrt{\frac{1-s^2}{1+s^2}}A_1 \right]\epsilon,\\
\frac{ds}{dt}&=\left[\half s\left(1-s^2\right)\beta_1 A_1\right]\epsilon^2.\
\end{eqnarray}

Similar to the results near \Rzero, this analysis reveals a spiraling motion around the \Rpi~manifold, but here the radius diverges exponentially. 
In contrast to the previous case, three distinct time scales are present: the derivative of $\theta$ is an order of magnitude larger than the derivative of $r$ and two orders of magnitude larger than the derivative of $s$. 
This means that the rotation around the \Rpi~manifold and the radial divergence away from the manifold occur more quickly than the translation along the manifold. In other words, each trajectory (and consequently any basin boundary) winds around \Rpi~within a plane with approximately fixed $s$ (see cross section in Figure~\ref{figure3} and Supplementary Video 3.
Rotation around the manifold occurs at a faster rate than divergence away, and both occur faster than translation along the manifold.

\section{Shape of the basin boundaries}
We can understand two qualitative aspects of the basin boundaries seen in numerics: 
(1) the basin boundaries are linear in the $s_c$-plane near \Rzero, and 
(2) the basin boundaries have spiral shape near \Rpi. 

The invariant ray \Rzero\, is surrounded by the \SSzero\, basin, and as we move further away form \Rzero, we enter the SD and DS basins, respectively  (Figs.~\ref{figure3} and \ref{SI_figure3}, \ref{fig:cd_coordinates}C).
The boundaries (separatrices) between \SSzero/SD and \SSzero/DS basins are the two stable manifolds leading to the SD/DS chimera saddles.
The relative width of the \SSzero\, and SD/DS basins varies with parameters $A$ and $\beta$ (Fig.~\ref{SIfigure1}), and the \SSzero\, basin is thin close to the origin, and for smaller $(A,\beta)$.

To understand the basin shapes near \Rzero, it is helpful to consider trajectories generated by points along a straight line orthogonal to \Rzero.  We will thereby use the asymptotic results~\eqref{eq:r_R0}-\eqref{eq:s_R0} derived in the previous section  valid for smaller $A$ and $\beta$.
Consider the set of points along a line segment parameterized by $k$ with $|k|\leq 1$ where $r=kr_0$, $\theta=\theta_0$ and $s=s_c$.  The trajectories generated by integrating the equations with initial conditions along that line segment will intersect the plane perpendicular to \Rzero~at some later time at $s= s_c+\delta$ where $\delta\ll\epsilon$.  By symmetry, if a point along the line with $k>0$ evolves toward the DS chimera, then the corresponding point with $k<0$ will be mapped to the SD chimera. Similarly, if a point with $k>0$ is mapped to the synchronized state, the corresponding point with $k<0$ will also be mapped to the synchronized state.    
\begin{figure*}[htp!]
 \centering
 \includegraphics[width=.9\textwidth]{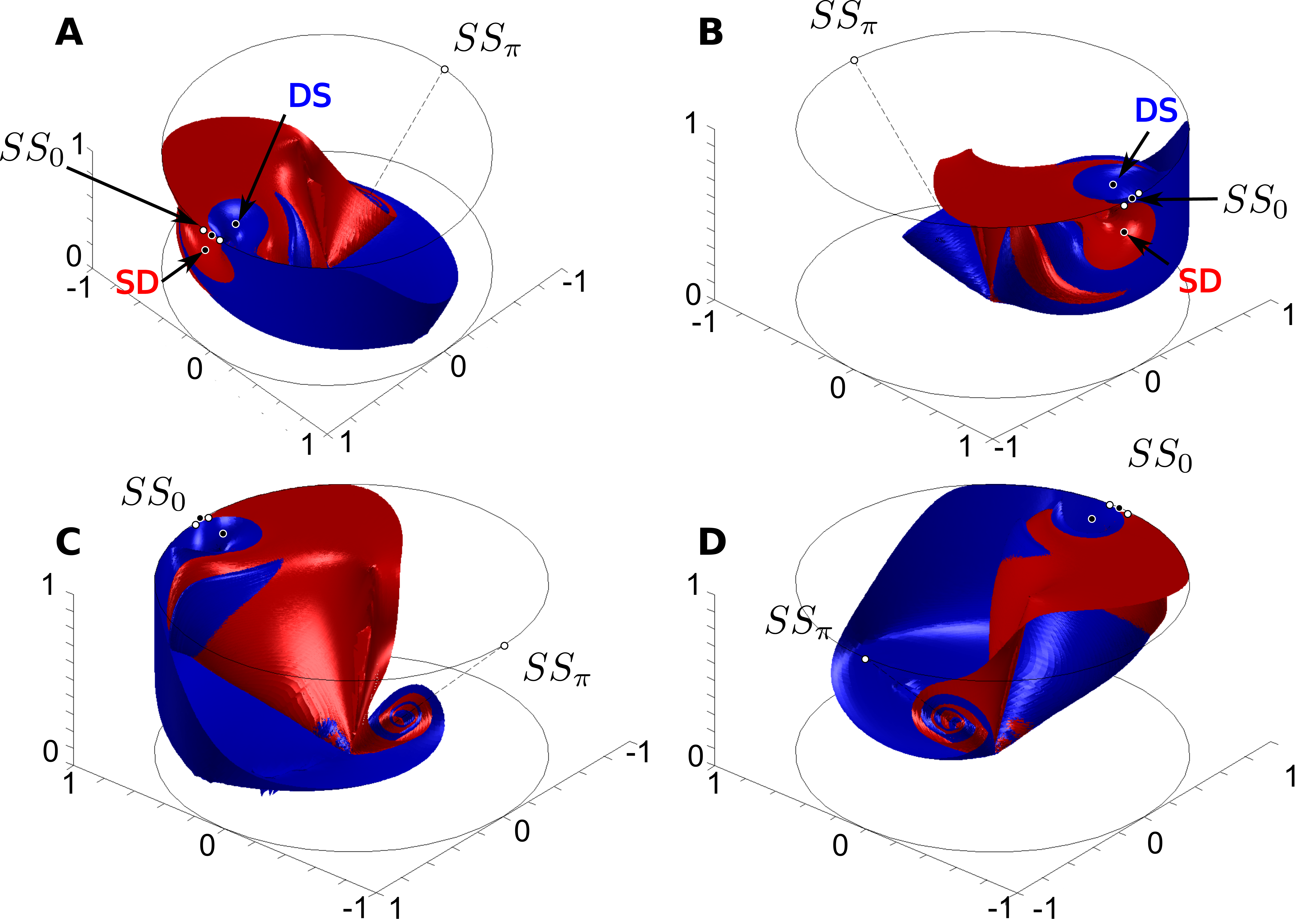}
 \caption{\label{SI_Separatrices1}
 Separatrix surfaces continued from the SD and DS saddle points on the $S_1$ (red) and $S_2$ (blue) manifolds, respectively, shown from different view angles {\bf (A, B, C, D)}. Continuation is performed as described in the text for $A=0.1$ and $\beta=0.025$. Black and white dots denote stable and unstable fixed points, respectively.
 }
\end{figure*}
Suppose $r_0\ll\delta$, so that all points along the line segment are chosen to be arbitrarily close to the \Rzero~manifold.  According to Eqs.~\eqref{eq:r_R0}-\eqref{eq:s_R0}, $d\theta$ is independent of $r$, so the images of these points in the plane $s_c+\delta$ will remain collinear. Now $dr=\bigO(\delta^2)$ and $ds=\delta$, so not only will the points along this segment remain collinear; as $s$ increases, they will also remain arbitrarily close to each other and to \Rzero. 
For at least one particular choice of $\theta_0$, this line segment will reach the surface of the cylinder in a direction tangent to the intersection of the invariant surfaces $S_1$ and $S_2$. As discussed above, this intersection is itself an invariant manifold, and thus the entire line segment will be mapped to the synchronized state, being the only attractor on the manifold. 
Thus there exists a line segment in the $s_c$ plane that lies in the basin of attraction for the $SS_0$ state and that separates the basins for SD and DS chimeras.  This suggests that, at least sufficiently close to \Rzero, the basin boundary between SD and DS chimeras must be linear.


Near \Rpi~the picture is different.  Because there are three distinct time scales, with the evolution of both $\theta$ and $r$ faster than the evolution of $s$, trajectories initially close to \Rpi~generate spirals within a fixed plane perpendicular to $s$ --- this is the origin of the spiral shape of the basin structure near \Rpi.

These qualitative arguments can be made rigorous in the limit where the \SSzero~basin becomes a set of measure zero (infinitesimal thickness). 
The basin boundaries (separatrices) are visualized in Fig.~\ref{SI_Separatrices1} (also see Fig.~\ref{SI_figure3} for a close up of the basins near \Rzero).

\section{Control strategies}

Determination of the structure of the basins of attraction for this system naturally invites the question of whether we can ``control'' the system. This can mean several things, among them: (1) can we intervene during an initial transient so as to direct the system to a desired equilibrium; (2) can we perturb the system to move it from one stable equilibrium to another; (3) can we stabilize an unstable equilibrium.  Each of these questions can also be examined with the goal of finding an ``optimal'' strategy of some kind, where optimality is usually defined as minimizing some aspect of the intervention.  To answer questions (1) and (2), knowledge of the basin structure in the thermodynamic limit $N \to \infty$ is clearly useful, at least for sufficiently large $N^{\sigma}$.

A full exploration of control and intervention strategies is beyond the scope of this paper, but as a demonstration of the power of our approach, we have performed a simple experiment. We wish to take a system at equilibrium in the DS chimera state, and to perturb it sufficiently that it goes to a different equilibrium (either SD or SS$_0$). We restrict ourselves to finite perturbations of the form $\theta_k^{(2)} \mapsto \theta_k^{(2)} + Q,\,k=1,\ldots,N^\sigma$, where $Q$ quantifies a uniform phase shift to all oscillators in the synchronous group (population 2). Since the perturbation leaves the system state on the invariant DS manifold, the final state may change from $DS$ to $SS_0$.

In the thermodynamic limit, this is equivalent to holding $\rho_1$ and $\rho_2=1$ constant while perturbing $\psi$ via $\psi \mapsto \psi - Q$.  Thus we expect the minimal required perturbation $Q_{\rm min}$ to be determined by the size of the restricted DS basin of attraction (restricted to the surface $\rho_2=1$, the top surface of the cylinder shown in Figure~2).  

\begin{figure}[htp!]
 \centering
 \includegraphics[width=0.6\columnwidth]{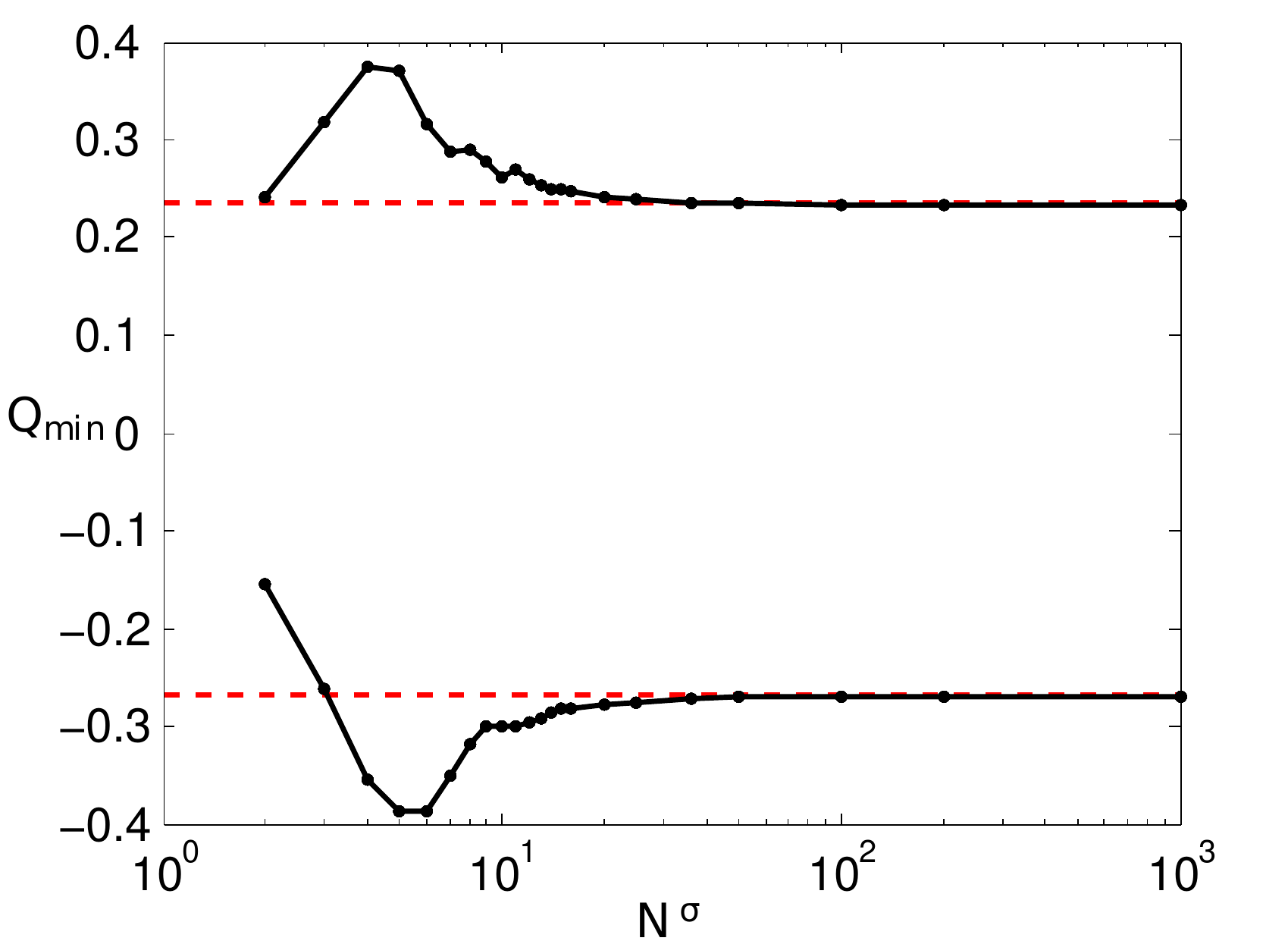}
 \caption{Minimal uniform perturbation of synchronous oscillator phases $Q_{\rm min}$ needed to escape from DS chimera state equilibrium.  Dashed line indicates asymptotic value for $N^{\sigma} \to \infty$.  Dots indicate results from numerical experiments at each fixed $N^{\sigma}$ value.}
 \label{SIfig:Qmin}
\end{figure}

Figure \ref{SIfig:Qmin} shows the expected asymptotic value of $Q_{\rm min}$ determined from our study of the basins of attraction presented here, as well as the $Q_{\rm min}$ values determined via numerical experiment for finite values of $N^\sigma$. The precise threshold varies slightly depending on the initial phases but asymptotes to the value observed in the continuum limit.  This good agreement confirms that (1) our knowledge of the basins of attraction in the thermodynamic limit can indeed inform control strategies for chimera states, and (2) insight into the finite system is indeed gained by analysis of the thermodynamic limit. 

Our analysis of the thermodynamic limit suggests that switching from the SS$_0$ state to a DS (or SD) chimera requires a perturbation to $\rho_1$ (or $\rho_2$), because $\rho_1=\rho_2=1$ is an invariant manifold. Hence, while a uniform phase shift ($\theta_i^{(2)} \mapsto \theta_i^{(2)} + Q)$ will perturb $\psi$, it will not be sufficient to desynchronize one of the two populations. Instead, a nonuniform phase shift that decreases the value of $\rho_1$ (or $\rho_2$) can accomplish the desired switching behavior.

Finally, we note that the control strategy presented here is quite naive.  It is likely that more ``optimal'' strategies exist in the sense that the perturbation magnitudes could be reduced.

\section{Discussion}

The probability that a random initial condition evolves to a chimera state, while important for real-world applications, has not been a frequent topic of investigation~\cite{Tinsley2012,Feng2015}.  Here, we have provided a detailed mathematical analysis unveiling the basin structure for a very simple system with two populations, allowing for insight into the chimera's relative rarity.
It remains to be seen whether similar efforts applied to other models such as neuronal or pulse-coupled oscillators (where reduction methods have become available only very recently~\cite{Pazo2014, Laing2014a}) will bear fruit, and how basin structure in those systems will compare. 

Oscillators on a ring with finite-range coupling exhibit chimera states that are (very long) transients with chaotic dynamics~\cite{Wolfrum2011b}. However, extensive computational analysis of the finite system~\eqref{eq:goveqns1} displays no such transient behavior while chaotic behavior is absent~\cite{OlmiMartens2015,Olmi2015}; this difference in dynamic behavior (along with others) seems to be related to differing coupling topology~\cite{OlmiMartens2015}. Moreover, very small oscillator systems with $N^\sigma=2,3,4$ are shown to display asymptotic stability of chimera states~\cite{Panaggio2015}.

Sampling the immense initial state space associated with the case of  $N^\sigma<\infty$ would be a burdensome task.  Our analysis was facilitated by considering  $N^\sigma\rightarrow\infty$, allowing us to focus on the low-dimensional order parameter dynamics on the OA-manifold~\cite{Ott2008b}.   While the higher dimensional dynamics off this manifold poses a challenge in its own right~\cite{Pikovsky2008}, the continuum theory allows us to gain useful insight by mapping the discrete to the continuous order parameter, $R_\sigma e^{i\Phi_\sigma}\approx\rho_\sigma e^{i\phi_\sigma}$ (identity for $N^\sigma\rightarrow\infty$).  Though bifurcation boundaries may blur for very low $N^\sigma$ and the finely filigreed basin boundaries near $R_\pi$ may break down, general basin structures will look similar even for moderate $N^\sigma$.

The stable manifolds of the saddles near \SSzero~divide phase space into simply connected basins of attraction (see Figs.~\ref{SI_Separatrices1}, \ref{SI_Separatrices2} and Supplementary Video 4). Basins near the fully synchronized (\SSzero) and chimera (SD/DS) states are simple in structure and relatively large\footnote{Local basin volumes of chimeras presumably scale like $|\mathbf{x}_\text{\tiny SD,DS}-\mathbf{x}_\text{\tiny SADDLE}|$, where $\mathbf{x}\in\mathbf{R}^3$ denotes coordinates of stable (SD/DS) and unstable (SADDLE) chimera states in the reduced phase space.}, resulting in robustness to perturbations (see also Fig.~\ref{SIfig:Qmin}). The twisting motion around the invariant rays $R_{0,\pi}$, however, yields a complex basin structure which we explain analytically. As one approaches $R_\pi$, the basin density diverges, and basins become \emph{locally} intermingled~\cite{Ott1994}: perturbations in that region affect the fate of a trajectory drastically. 

Continuum theory allows us to construct initial phase densities leading to chimera states via $f^\sigma(\theta,t)=\frac{1}{2\pi}\left [1+\sum_{n=1}^\infty \rho_\sigma^n e^{in(\theta-\phi_\sigma)} +c.c.\right]$ (see \ref{app:OA}). If instead initial phases $\theta_k$ are sampled uniformly on $[-\pi,\pi]$, the probability distribution for $R_\sigma$ is unimodal with mean $\sim 1/\sqrt{N^\sigma}$ and variance $\sim 1/N^\sigma$.  The probability distribution for $\Phi_\sigma$ is invariant and uniform; this observation combined with the intermingled basin structure near \Rpi~explains why in practice random initial phases can lead to both chimera and fully synchronized states. Thus, this implies in particular that chimera states, given random initial conditions, are not always rare occurrences (depending on parameter values, see also Fig.~\ref{SIfigure3}).


The results we have presented focus on the case of two populations. 
While two populations allow for multi-stability between the fully synchronous state (\SSzero) and two symmetrically equivalent chimera states (SD and DS),  generalizations of such hierarchical structure to $n>2$ populations~\cite{Martens2010var,Shanahan2010} make accessible larger configuration spaces of size $2^n$ by variation of the synchronization-desynchronization patterns. 
One may wonder if any of the structures studied here retains relevance in cases of more than two populations~\cite{Martens2010bistable,Shanahan2010}. It can be shown that the invariant hyper-ray corresponding to $R_0$ defined by $\rho_\sigma=\rho$ and $\phi_\sigma=0$ exists for $n>2$, and that the flow on this ray is $\rho\rightarrow 1$. This suggests that the phase space is skeletonized similarly and a similar analysis may be feasible---a task left for a future study.

Biologically, systems with multiple coexisting chimera attractors have been proposed to describe `metastable dynamics' required to modulate neural activity patterns~\cite{TognoliKelso2014}, or to encode memory. Indeed, localized dynamical states are directly related to function in neural networks~\cite{Fell2011,Hubel1959}; 
localized synchrony has been widely studied in neural field models as bump states~\cite{Amari1977,Laing2001a,Laing2014a}, and are phenomenologically similar to chimera states.  
It is worth noting that chimeras occur in models of neural activity~\cite{Sakaguchi2006, Olmi2010, Cabral2011, Hizanidis2013, Vullings2014}.

Implementations of chimera states could likely be achieved in micro-(opto)-electro-mechanical oscillators~\cite{Eichenfield2009, Heinrich2011} where synchronization patterns may have technological applications.  Conversely, as power grid network topologies evolve to incorporate growing sources of renewable power, the resulting decentralized, hierarchical networks~\cite{Rohden2012, Motter2013} may be threatened by chimera states, which could lead to large scale partial blackouts and unexpected behavior. 

The potential for applications---or threats---makes the dynamic re-configuration and switching between chimera configurations (possibly modulating functional properties of the underlying oscillator network) particularly relevant~\cite{BickMartens2014}; applications that modulate functional properties can only be achieved using detailed knowledge of the basin structure. 
As a test of principle, we successfully implemented a simple  algorithm to move the finite oscillator system between different equilibria, demonstrating that an understanding of the basins of attraction for the $N^{\sigma} \to \infty$ system has value.  We hope that future work will explore the construction of efficient control strategies to stabilize or prevent chimera states, with applications across many fields.



\ack
Research supported by the Dynamical Systems Interdisciplinary Network, University of Copenhagen (EAM). We thank C.~Bick, R.~Mirollo, S.~Strogatz and O.~Omel'chenko for valuable conversations and comments, and anonymous referees for helpful suggestions to improve the manuscript.

\appendix
\section{Derivation of OA-reduced equations}\label{app:OA}

We consider the Kuramoto-Sakaguchi model with non-local coupling between $n$ populations~\cite{Abrams2008,Montbrio2004}
\begin{eqnarray}\label{eq:goveqns11}
 \dot{\theta_k^{\sigma}} &=& 
 \omega + \sum_{\sigma'=1}^n \frac{K_{\sigma\sigma'}}{N^{\sigma'}}\sum_{l=1}^{N^{\sigma'}}
 \sin{(\theta_l^{\sigma'}-\theta_k^\sigma-\alpha)},\
\end{eqnarray}
where $\theta_k^{\sigma}$ is the phase of the $k$th oscillator $k= 1, \ldots, N^{\sigma}$ belonging to population $\sigma = 1, \ldots, n$.
To facilitate comparison with previous work~\cite{Abrams2008,Montbrio2004,Martens2010var}, we consider the case of symmetric coupling with $K_{\sigma\sigma'}=K_{\sigma'\sigma}$. The phase lag parameter $\alpha$ tunes between the regimes of pure sine-coupling ($\alpha=0$) and pure cosine-coupling ($\alpha=\pi/2$). 
In what follows, we introduce the re-parameterized phase lag parameter $\beta=\pi/2-\alpha$, since for this type of system chimeras emerge in the limit of cosine-coupling~\cite{Abrams2008,Martens2010var}, i.e. $\beta\rightarrow0$.

To make further progress, we consider the thermodynamic limit, i.e., the case of $N^\sigma\rightarrow \infty$ oscillators per population. This allows for a description of the dynamics in terms of the mean-field order parameter~\cite{Ott2008b,Ott2009,Ott2011}. 
Eqs.~\eqref{eq:goveqns11} then give rise to the continuity equation
\begin{eqnarray}
 \frac{\partial f^{\sigma}}{\partial t} + \frac{\partial}{\partial\theta}(f^\sigma v^\sigma)=0,\
\end{eqnarray}
where $f^\sigma(\theta,t)$ is the probability density of oscillators in population $\sigma$, and $v^\sigma(\theta,t)$ is their velocity, given by
\begin{eqnarray}
 v^\sigma(\theta,t) &=& \omega +\sum_{\sigma=1}^n K_{\sigma\sigma'}\int e^{i\theta'}f^{\sigma'}(\theta',t)d\theta'.\
\end{eqnarray}
Here we have dropped the superscripts to simplify notation: $\theta$ means $\theta^\sigma$, and $\theta'$ means $\theta^{\sigma'}$.
Following Ott and Antonsen~\cite{Ott2008b,Ott2009}, we consider probability densities along a manifold given by 
\begin{eqnarray}
f^\sigma(\theta,t)=\frac{1}{2\pi}+\left\{\frac{1}{2\pi}\sum_{n=1}^\infty \left[a^*_{\sigma}(t)e^{i\theta}\right]^n +c.c.\right \}
\end{eqnarray}
where $^*$ denotes complex conjugation and $a_{\sigma}(t)$ is given by 
\begin{eqnarray}
a_{\sigma}(t)=\int e^{i\theta}f^{\sigma}(\theta,t)d\theta.
\end{eqnarray}
Defining $a_{\sigma}(t)=\rho_\sigma(t)e^{i\phi_\sigma(t)}$ where $\rho_\sigma$ and $\phi_\sigma$ represent mean-field order parameters, the governing equation can be reduced to the system  \cite{Abrams2008,Martens2010var,Panaggio2015_2}
\begin{eqnarray}\label{eq:OPrho}
 \dot{\rho}_\sigma&=&\frac{1-\rho_\sigma^2}{2}\sum_{\sigma'=1}^n K_{\sigma\sigma'} \rho_{\sigma'} \sin{(\phi_{\sigma'}-\phi_\sigma+\beta)}\\\label{eq:OPphi}
 \dot{\phi}_\sigma&=&\omega-\frac{1+\rho_\sigma^2}{2\rho_\sigma}\sum_{\sigma'=1}^n K_{\sigma\sigma'} \rho_{\sigma'} \cos{(\phi_{\sigma'}-\phi_\sigma+\beta)}.\
\end{eqnarray}
The Ott/Antonsen manifold, in which the Fourier coefficients $f_n(t)$ of the probability density $f$ satisfy $f_{n}(t)=a(t)^n$, is globally attracting for a frequency distribution with non-zero width $\Delta$~\cite{Ott2009}. For identical oscillators ($\Delta=0$), the dynamics for the problem (with $n=2$ populations) can be described by reduced equations using the Watanabe/Strogatz ansatz~\cite{WatanabeStrogatz1994}, as shown in Pikovsky and Rosenblum~\cite{Pikovsky2008}; the authors showed that Eqs.~\eqref{eq:goveqns1} may also be subject to more complicated dynamics than those described by the Ott/Antonsen ansatz. Studies by Laing~\cite{Laing2009,Laing2012b} investigated the dynamics using the Ott/Antonsen ansatz for $n=2$ populations for the case of non-identical frequencies and found that the dynamics for sufficiently small $\Delta$ is qualitatively equivalent to the dynamics obtained for $\Delta=0$. It is therefore justified to discuss the dynamics for $\Delta\rightarrow 0$ representing the case of \emph{nearly} identical oscillators using the Ott/Antonsen reduction.

\section{Governing equations for two populations}\label{app:goveqns2pop}
We restrict our attention to the case of $n=2$ populations. Accordingly, we define the coupling parameters
$K_{11}=K_{22}=\mu$ and $K_{12}=K_{21}=\nu$; by rescaling time we can eliminate one parameter so that $1=\mu+\nu$ without loss of generality. The remaining parameter is redefined via $A=\mu-\nu$, expressing the disparity of coupling between the two neighboring populations. By virtue of the translational symmetry, $\phi_\sigma\rightarrow\phi_\sigma+\text{const.}$, the dynamics of the system is effectively three dimensional.  We introduce the angular phase difference $\psi=\phi_1-\phi_2$ of the order parameter, and the resulting governing equations become
\begin{eqnarray}\label{eq:rho1rho2psi_eqnA}
 \dot{\rho}_1 &=& \frac{1-\rho_1^2}{2}\left[\mu \rho_1\sin{\beta} + \nu \rho_2 \sin{(\beta-\psi)} \right],\\
 \dot{\rho}_2 &=& \frac{1-\rho_2^2}{2}\left[\mu \rho_2\sin{\beta}+\nu \rho_1\sin{(\beta+\psi)}  \right],\\\nonumber
 \dot{\psi} &=& \frac{1+\rho_2^2}{2\rho_2}\left[\mu \rho_2\cos{\beta} + \nu\rho_1\cos{(\beta+\psi)} \right]\\\label{eq:rho1rho2psi_eqnC}
 &-& \frac{1+\rho_1^2}{2\rho_1}\left[\mu\rho_1 \cos{\beta} + \nu\rho_2\cos{(\beta-\psi)} \right].\
\end{eqnarray}
where the state variables lie in the domain $D=\{(\rho_1,\rho_2,\psi)\in \mathbb{R}^3 | 0<\rho_1,\rho_2 \leq 1,\psi \in [-\pi,\pi]\}$.

To investigate the basins of attraction, it proves useful to express the dynamics in terms of the sums and difference of the order parameters, i.e., we define
\begin{eqnarray}\label{eq:defsd}
s &=& \frac{1}{2}(\rho_1 + \rho_2), \\
d &=& \frac{1}{2}(\rho_1 - \rho_2), \
\end{eqnarray}
and $\psi$ as above.
These variables belong to the domain defined by $\psi\in[-\pi,\pi]$, $s\in[0,1]$ and $d\in[-a,a]$ with $a(s)=\frac{1}{2}-|\frac{1}{2}-s|$
(the back-transformation is $\rho_1=s+d$ and $\rho_2=s-d$, without the factor of 2.).
The governing equations are then expressed as
\begin{eqnarray}\nonumber
\dot{s}&=& \frac{1}{2}\{s [\mu (1-3 d^2-s^2)  + \nu(1+d^2-s^2)   \cos{\psi}] \sin{\beta} \\\label{eq:sdpsi_eqnA}
       &&+ \nu d(1-d^2+s^2) \cos{\beta} \sin{\psi}\}, \\\nonumber
\dot{d}&=& \frac{1}{2}\{d [\mu (1-d^2-3 s^2)  - \nu(1-d^2+s^2)   \cos{\psi}] \sin{\beta} \\ 
       &&- \nu s(1+d^2-s^2) \cos{\beta} \sin{\psi}\}, \\\nonumber
\dot{\psi}&=&(d^2-s^2)^{-1}\big\{-2 d s \cos{\beta} [\mu(d^2-s^2)  + \nu \cos{\psi}] \\\label{eq:sdpsi_eqnC}
&&+ [s^2 + d^4 + s^4 +d^2 (1 - 2 s^2)] \nu \sin{\beta} \sin{\psi}\big\}.\
\end{eqnarray}

Eqs.~\eqref{eq:rho1rho2psi_eqnA}-\eqref{eq:rho1rho2psi_eqnC} or Eqs.~\eqref{eq:sdpsi_eqnA}-\eqref{eq:sdpsi_eqnC}, respectively, are invariant under the transformation $\Sigma: (\rho_1,\rho_2,\psi)\mapsto(\rho_2,\rho_1,-\psi)$, corresponding to interchanging the two oscillator populations. 
More generally, the change of parameters $\Pi:\beta \mapsto \beta+\pi$ reverses time in the governing equations, thus inverting flow and stability properties in phase space. This is also valid for the more general case of $n>2$ equations,~i.e., for Eqs.~\eqref{eq:OPrho} and \eqref{eq:OPphi}.

\section{Invariant manifolds (IMs).}\label{app:IM}

\paragraph{Two-dimensional invariant manifolds.} Letting $\rho_1\rightarrow 1$ in Eqs.~\eqref{eq:rho1rho2psi_eqnA}-\eqref{eq:rho1rho2psi_eqnC} leaves $\rho_1$ invariant, i.e. $\dot{\rho}_1=0$. The same holds true for $\rho_2$ by symmetry.  Thus we find two two-dimensional invariant surfaces,  corresponding to the top and side surface of $D$, defined by
\begin{eqnarray}\label{eq:M01}
 S_\sigma &=& \{(\rho_1,\rho_2,\Psi) \in S \;|\; \rho_\sigma=1\},\ 
\end{eqnarray}
where $\sigma={1,2}$ refers to the SD, DS manifolds, respectively. 
The dynamics in one manifold is identical to the other via the symmetry operation defined by operator $\Sigma$, see main text.
The dynamics on these IMs is analyzed in~\cite{Abrams2008}.

\paragraph{One-dimensional invariant manifolds.}
Our numerical investigations indicate the presence of an invariant manifold at $\psi=0,\pi$ with 
$\rho_1=\rho_2$. Substituting these values into Eqs.~\eqref{eq:rho1rho2psi_eqnA}-\eqref{eq:rho1rho2psi_eqnC}, we get
\begin{eqnarray}\label{eq:MphiEqns}
\dot{s}&=& \frac{1}{2}\sin{\beta}\cdot s  (1-s^2) , \\
\dot{d}&=& \dot{\psi} =0 .\
\end{eqnarray}
The first equation implies that any initial point on the rays with $d=0$ and $\psi=0,\pi$ remains there for all times; if $0<\beta<\pi$, the trajectory moves towards the SS$_0$ attractor according to \eqref{eq:MphiEqns}. Thus, two invariant rays exist, defined via 
\begin{eqnarray}\label{eq:Mphi}
R_\phi &=& \{(\rho_1,\rho_2,\Psi) \in D| \rho_1=\rho_2 \text{ and } \psi=\phi\}\ 
\end{eqnarray}
with $\phi = 0,\pi$.

Note that another one-dimensional invariant manifold $S_{12}$ is defined as the intersection $S_1\cap S_2$, and any initial point with $s=1$ on $S_{12}$ will therefore always end up in the SS$_0$ state.

\section{Fixed points.}\label{app:fixedpoints}
\paragraph{Fixed points on \texorpdfstring{$S_{1,2}$}{S1,S2}.}
The fixed points in the $S_{1,2}$ manifolds are the SD, DS chimera states and fully synchronized SS$_0$ states that are discussed in detail in~\cite{Abrams2008}; note that since $S_{12}$ is an invariant manifold, there must be another fixed point in addition to SS$_0$ contained in it, with opposite stability:
this source is found at $(\rho_1,\rho_2,\psi)=(1,1,\pi)$, which we refer to as $SS_\pi$. Figure 2 illustrates how trajectories nearby are repelled from the ray $R_\pi$. 
On $S_{1,2}$, stable chimera states are born through a saddle node bifurcation, and undergo a Hopf bifurcation for sufficiently large disparity values $A$ so that $\rho_\sigma<1$ is oscillatory; the associated limit cycle is destroyed in a homoclinic bifurcation with even larger $A$.

\paragraph{Chimera states.}
In addition to the in-phase ($\rho_1=\rho_2=1$ and $\psi=0$) and anti-phase ($\rho_1=\rho_2=1$ and $\psi=\pi$) equilibrium points,  there are also three equilibrium points with $\rho_2=1$ and $\rho_1\neq 1$ (and three analogous fixed points with $\rho_1=1$ and $\rho_2\neq 1$)~\cite{Panaggio2015_2}.  These equilibrium points represent chimera states.  Numerics suggest that two of these equilibrium points occur near $\psi=0$ and one occurs near $\psi=\pi$. Using an ansatz motivated by these numerical results, we find that these equilibrium points satisfy the following scaling relationships (where $A=\mu-\nu$ and $\mu+\nu=1$:

\subparagraph{i) Stable Chimera near \texorpdfstring{$\psi=0$}{psi=0} (DS):}
\begin{eqnarray*}
\beta  & \sim & A \beta_1,\\
\rho_1 & \sim & 1 - A \left(1+\sqrt{1-4\beta_1^2}\right) + A^2 \left( 1-4\beta_1^2 + \frac{1-6\beta_1^2}{\sqrt{1-4\beta_1^2}} \right),\\
\psi   & \sim & A\left(2\beta_1\right)+A^2\left(\beta_1\left[1-\sqrt{1-4\beta_1^2}\right]\right).
\end{eqnarray*}

\subparagraph{ii) Unstable Saddle Chimera near \texorpdfstring{$\psi=0$}{psi=0} (UC):}
\begin{eqnarray*}
\beta  & \sim A & \beta_1,\\
\rho_1 & \sim & 1 - A \left(1-\sqrt{1-4\beta_1^2}\right) + A^2 \left( 1 - 4 \beta_1^2 - \frac{1-6\beta_1^2}{\sqrt{1-4\beta_1^2}} \right),\\
\psi&\sim & A\left(2\beta_1\right)+A^2\left(\beta_1\left(1+\sqrt{1-4\beta_1^2}\right)\right).\
\end{eqnarray*}

\subparagraph{iii) Unstable Chimera near \texorpdfstring{$\psi=\pi$}{psi=pi} (UC):}
\begin{eqnarray*}
\beta&\sim A^{\frac{1}{2}}\beta_1,\\
\rho_1&\sim 1-A^2\left(2\beta_1^2\right)+A^3\left(\frac{2}{3}\beta_1^4\right),\\\
\psi&\sim\pi-A^{\frac{3}{2}}\left(2\beta_1\right)+A^{\frac{5}{2}}\left(\frac{4}{3}\beta_1^3-2\beta_1\right).
\end{eqnarray*}

These relationships are useful when trying to solve for the precise fixed point locations numerically and for approximating their stable and unstable manifolds in order to deduce the basin boundaries. We note that chimera states asymptotically approach either \SSzero~or \SSpi~
as $(A,\beta)\rightarrow (0,0)$.

\paragraph{Origin.} 
The origin is an unstable fixed point, as can be seen by linearizing for small $\rho_1$ and $\rho_2$ in Eqs.~\eqref{eq:rho1rho2psi_eqnA}--\eqref{eq:rho1rho2psi_eqnC}.

\paragraph{Other fixed points. }
Here we ask whether there are any fixed points off the invariant manifolds $S_{1,2}$: i.e., are there any fixed points with $\rho_1, \rho_2 \notin S_{1,2}$ where no population is completely synchronized?
Together with Eqs.~\eqref{eq:rho1rho2psi_eqnA}-\eqref{eq:rho1rho2psi_eqnC}, $0<\rho_1,\rho_2<1$ implies the following conditions:
\begin{eqnarray}
 0 &=& \mu\rho_1\sin{\beta} + \nu\rho_2\sin{(\beta-\psi)}, \\
 0 &=& \mu\rho_2\sin{\beta} + \nu \rho_1 \sin{(\beta+\psi)},  \\\nonumber
 0 &=& \rho_1(1+\rho_2^2)\left[\mu\rho_2\cos{\beta} + \nu\rho_1\cos{(\beta+\psi)}  \right]\\ 
   &-& \rho_2(1+\rho_1^2)\left[\mu\rho_1\cos{\beta} + \nu\rho_2\cos{(\beta-\psi)} \right].\
\end{eqnarray}
We know that $\beta\rightarrow \beta + \pi$ reverses time, so we can w.l.o.g. restrict our attention to $0\leq \beta \leq  \pi$.
When $\beta=0,\pi$, the first two equations are satisfied if $\psi=0,\pi$. The third equation yields the solutions $\rho_2=\pm\rho_1$ and $\rho_2= \frac{\rho_1(1+A) \pm\sqrt{-4 (-1+A)^2+(3-A) (3 A-1) \rho_1^2}}{2 (A-1)}$, where only the first branch lies in $0\leq \rho_{1,2}\leq 1$.

For all other cases, let us consider the equations by introducing $K=\mu/\nu>1$ and $\rhorel=\rho_2/\rho_1$:
\begin{eqnarray}
 0 &=& [\cos{(\psi)} \rhorel + K] \sin{\beta} - \rhorel \cos{\beta} \sin{\psi},\\
 0 &=& [K \rhorel + \cos{\psi}] \sin{\beta} + \cos{\beta}\sin{\psi},\\
 0 &=& -[2 \rhorel^2 \rho_1^3+(\rhorel^2-1)\rho_1]\sin{\psi}\sin{\beta} \label{thirdeqn} \\ 
 &&+ [(1-\rhorel^2)\rho_1\cos{\psi} + (\rhorel^2-1) K \rhorel \rho_1^3]\cos{\beta}. \nonumber
\end{eqnarray}
We note now that $\rhorel>0$ and $\sin{\beta}>0$ by assumption and we can eliminate these expressions as follows
\begin{eqnarray}\label{KR1}
 \frac{\cos{\beta}\,\sin{\psi}}{\sin{\beta}} &=& \frac{\cos{(\psi)}\,\rhorel + K}{\rhorel},\\\label{KR2}
 \frac{\cos{\beta}\,\sin{\psi}}{\sin{\beta}} &=& -K \rhorel - \cos{\psi}.\
\end{eqnarray}
Equating \eqref{KR1} and \eqref{KR2}, it follows that  a fixed point with $0<\rho_1,\rho_2<1$ and $0<\beta<\pi$ can only exist if 
\begin{eqnarray}
 0 &=& K \rhorel^2 + 2 \cos{(\psi)}\,\rhorel + K,\
\end{eqnarray}
which has the solutions
\begin{eqnarray}
 \rhorel &=& -\frac{\cos{\psi}\pm\sqrt{\cos^2{(\psi)}-K^2}}{K}.\
\end{eqnarray}
However, by assumption, we have $K>1$ and the solutions are complex. Therefore, even if we find real values $\psi,\rho_1$ that satisfy the third fixed point equation \eqref{thirdeqn}, there will be no real solutions for $\rhorel$, and thus also not for $\rho_2$.
Therefore fixed points in the interior of the domain can be excluded when $\beta\neq n\pi$ where $n$ is an integer.


\section{Numerical Analysis}\label{app:numerics}
\subsection{Probabilistic measure of basins of attraction}
In order to obtain an estimate of the sizes of the basins of attraction of the equilibria of \eqref{eq:rho1rho2psi_eqnA}-\eqref{eq:rho1rho2psi_eqnC}, we selected 1000 random initial points $(\rho_1,\rho_2,\psi)$. Eqs.~\eqref{eq:rho1rho2psi_eqnA}-\eqref{eq:rho1rho2psi_eqnC} were then integrated for a sweep of parameter values of $0.01\leq A \leq  0.49$ with increments of $0.01$ and $0.005\leq \beta \leq 0.245$ with increments of $0.005$ until a final state was detected. The contour plot in Figure~\ref{SIfigure3} displays the fraction of those trajectories with final states near a chimera state.

It should be noted that the numerical experiment above assumes that $\rho_1$, $\rho_2$, and $\psi$ are uniformly distributed. For systems with a finite number of oscillators $N^{\sigma}$ in each population, the expected value of the order parameter value $\rho_{\sigma}$ is $\mathcal{O}(1/\sqrt{N^\sigma})$. Hence, the probabilities computed using the above scheme should not be interpreted as the probability that a state with randomly selected initial phases $\theta_k^{(\sigma)}$ would evolve toward a chimera state. Instead, they represent the size of the basins of attraction of the chimera states relative to the size of the basin of attraction of the fully synchronized state in the continuum limit $N^\sigma\rightarrow \infty$.

\subsection{Destination Maps}
Simulations for a given initial condition were carried out until trajectories to a fully synchronized (limit point, LP) or a stable (LP) or breathing chimera (limit cycle LC) occurred. The detection of these three types of states was carried out in two steps, described below.
Integration of Eqs.~\eqref{eq:rho1rho2psi_eqnA}-\eqref{eq:rho1rho2psi_eqnC} or \eqref{eq:sdpsi_eqnA}-\eqref{eq:sdpsi_eqnC} were carried out in Matlab\texttrademark~using the \verb|ode45| solver routine 
with event detection (see below) on a high performance computation cluster, with a relative error tolerance of $10^{-8}$. Algorithms below are outlined for $(\rho_1,\rho_2,\psi)$-coordinates; analogous detections for $(s,d,\psi)$-coordinates are carried out by applying the related coordinate transformations.
Below, $d\rho_{1,2} \approx \dot{\rho}_{1,2}\,dt $ and $d\psi\approx \dot{\psi}\, dt$ denote the approximate differential values evaluated by the o.d.e. integrator at discrete time steps.

\subsection{Simple convergence test (Event A)}
This simple test was used to detect the type of state is asymptotically achieved.
Integration was stopped by an event detection algorithm solving for roots of:
\begin{itemize}
  \item LP detection: $v = [d\rho_1^2 + d\rho_2^2 + d\psi^2]^{1/2}-\delta$: convergence to any LP (in any direction).
  
  \item Convergence to LC on \Mone: $v = [(\rho_1-1)^2  + d\rho_2^2] - \delta $,
  passage through $\dot{\rho_2}=0$ (= 1 cycle), positive direction.
  
  \item Convergence to LC on \Mtwo: $ v =[ (\rho_2-1)^2 + d\rho_1^2]-\delta$,
  passage through $\dot{\rho_1}=0$ (= 1 cycle), positive direction.

\end{itemize}
A convergence tolerance  of $\delta \sim 10^{-6}$ was chosen.

\subsection{Estimating time to attractor}
The following algorithm is adopted for obtaining estimates for the time to reach the attractor, $T$, i.e., the traveling time from initial to end condition. When these times are not of interest, the previous scheme is preferred due to significant gains in computational speed. 
\begin{enumerate}
  \item Integration is carried out until $v = [d\rho_1^2 + d\rho_2^2 + d\psi^2]^{1/2}-\delta$ crosses a zero ({\bf Event B}, LP detection).
  
  \item If the integration fails to detect a fixed point, the algorithm enters a loop of max. 100 iterations, where:
  
  \subitem i.) Integration is carried out to detect $k=1,\ldots,10$ events of  type {\bf Event A}, limit point and limit cycles). Periods of limit cycles and event states $(\rho_1,\rho_2,\psi)|_{t=t_k}$ are stored.
  
  \subitem ii.) Test for convergence to limit point or limit cycle: 
  $||(\rho_1,\rho_2,\psi)|_{t=t_{k}}-(\rho_1,\rho_2,\psi)|_{t=t_{k+1}} ||< \epsilon_c$ with $\epsilon_c \sim 10^{-4}$
  \subitem iii.) Exit loop when a  LP or LC is detected or 100 iterations are carried out.
  
  \item If LC or LP is detected, the final state is detected as explained above. Otherwise, failed convergence is stored as a failed end state.
\end{enumerate}

%
%
%
\begin{figure*}[htp!]
 \centering
 \includegraphics[width=0.8\textwidth]{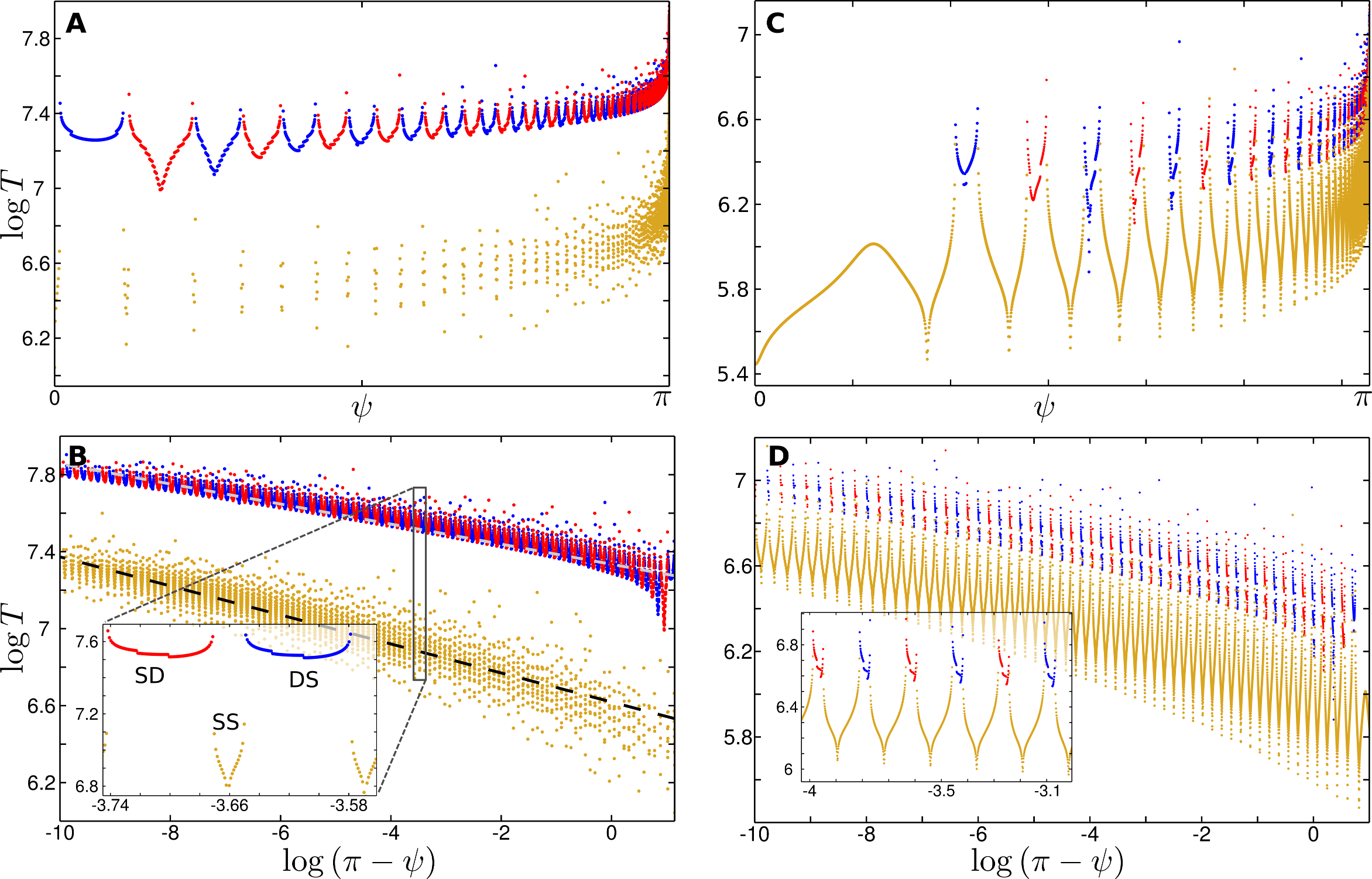}
 \caption{\label{SIfigure2} Times to attractor. Parameter values for ({\bf A, B}): $A=0.1, \beta = 0.025$ at $\rho_1=\rho_2 = 0.56625$ and for ({\bf C, D}): $A=0.1, \beta = 0.05$ at $\rho_1=\rho_2 = 0.5$. Final destinations are color coded  in red for SD, blue for DS and yellow for ${\rm SS}_0$ states.}
\end{figure*}
\subsection{Destination maps in the \texorpdfstring{$s_c$}{sc}-plane }
\begin{figure*}[htp!]
  \centering
  \includegraphics[width=0.8\textwidth]{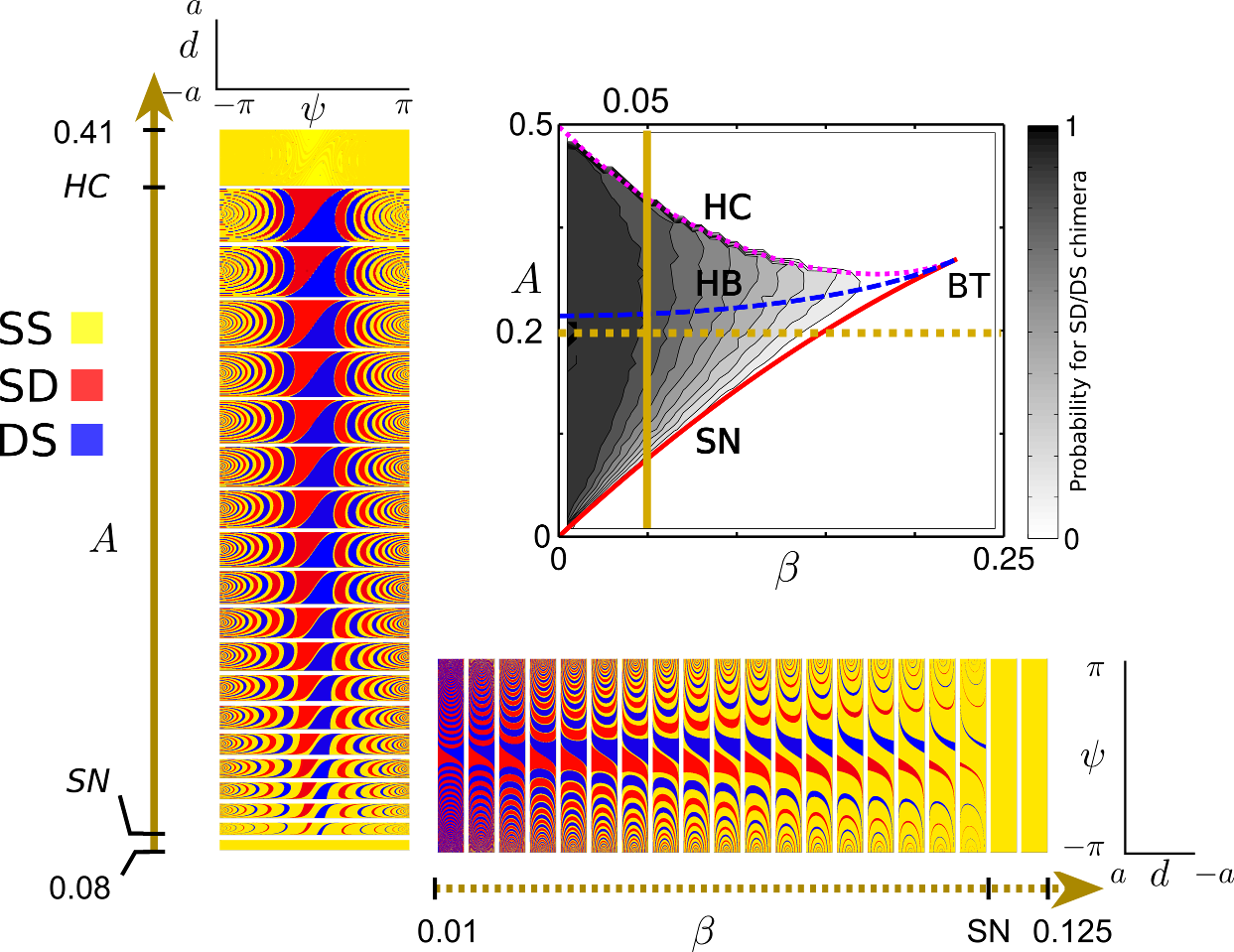}
  \caption{\label{SIfigure1} Destination maps for parameter sweeps in $A$ (vertical) and $\beta$ (horizontal), respectively. The maps are shown in the $(d,\psi)$-plane at $s=s_c=1-A$. Parameters where saddle node and homoclinic bifurcations occur are denoted by SN and HC, respectively. 
  }
\end{figure*}

Destination maps were calculated for $\beta=0.01,\ldots,0.125$ at constant $A=0.2$ (Figure~\ref{SIfigure1}A) and for $A=0.08,\ldots,0.41$ at constant $\beta=0.05$ (Figure~\ref{SIfigure1}B). 
Saddle-node (SN) and homoclinic (HC) transitions are indicated.

\subsection{Destination map for small \texorpdfstring{$s$}{s}}
\begin{figure}[htp!]
  \centering
\includegraphics[width=0.6\textwidth]{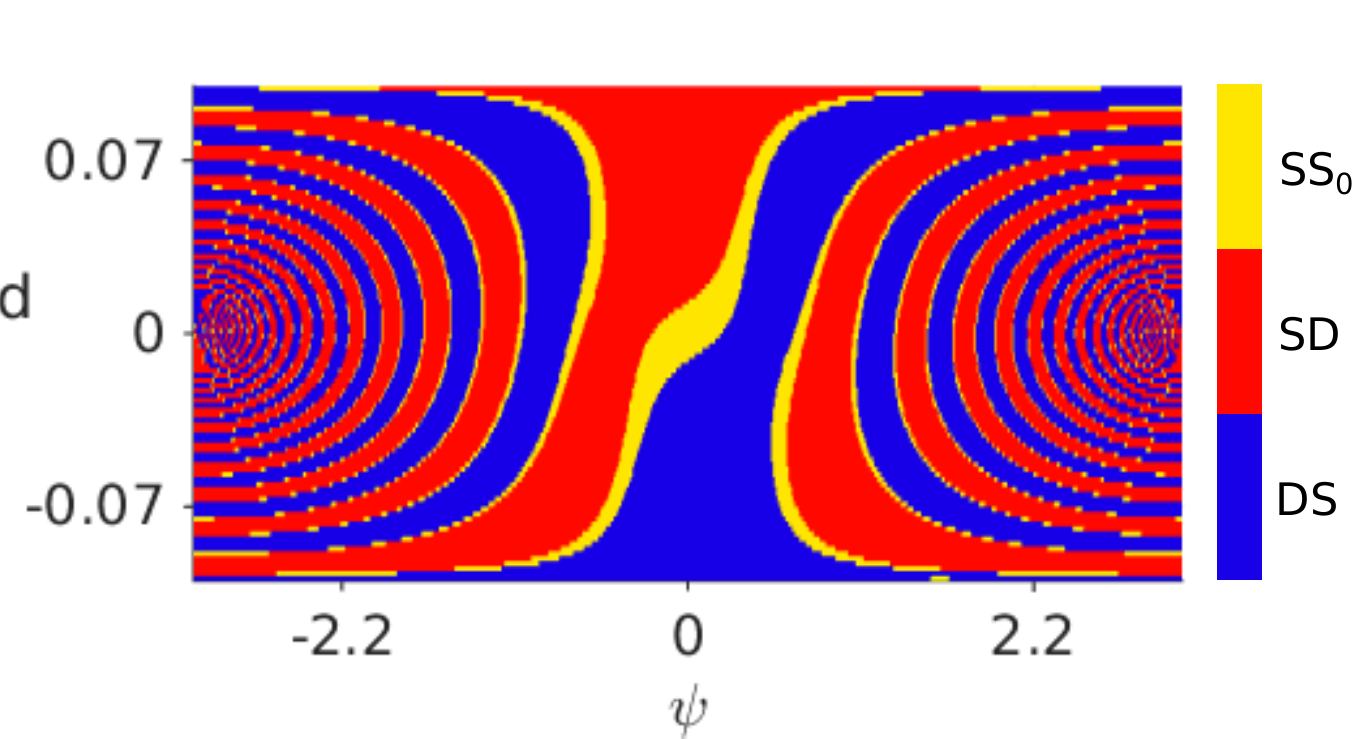}
  \caption{Destination map in the $(d,\psi)$ plane with $s=0.1$. Red indicates SD chimera, blue indicates DS chimera, yellow indicates SS$_0$ state. Parameters are $A=0.1$ and $\beta= 0.025$. }
  \label{SI_figure3}
\end{figure}

In Figure~\ref{SI_figure3}, we display a sample destination map computed for small $s$ for the purpose of demonstrating what the basin structure would look like if initial phases were chosen from a uniform random distribution.  If, for example, $N^\sigma=50$, the expected initial value of $s$ would be close to $0.1$.

\subsection{Numerical continuation of the basin boundaries (separatrices)}\label{app:numerics_cont}
\begin{figure*}[htp!]
 \centering
 \includegraphics[width=0.9\textwidth]{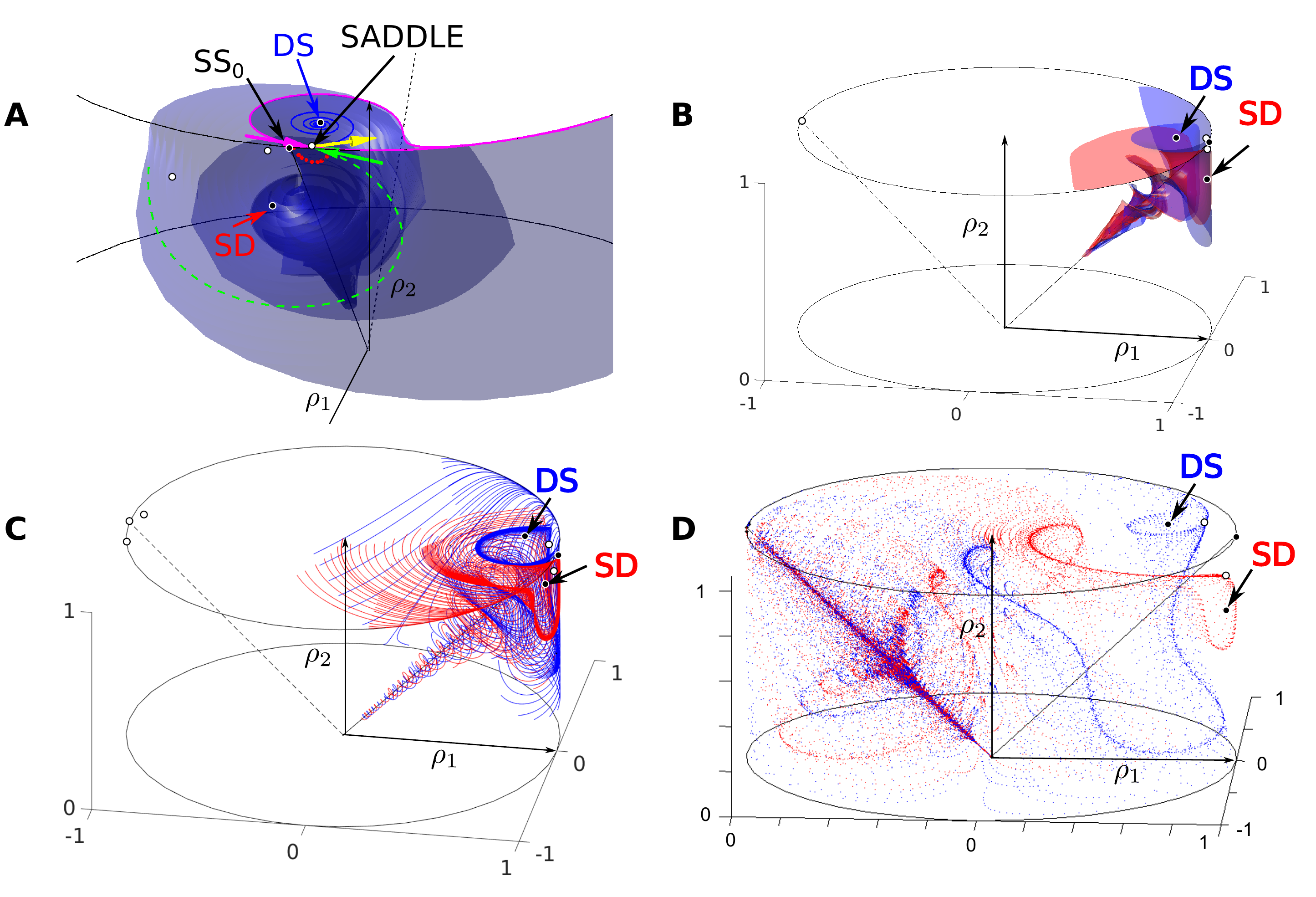}
 \caption{\label{SI_Separatrices2} 
 Visualization of separatrix surfaces and trajectories. Points along the separatrix corresponding the DS chimera state are colored blue and points along the separatrix corresponding the SD chimera state are colored red. 
 {\bf A:} Continuation of the separatrix for the DS chimera state. Stable manifold and corresponding eigenvectors of SADDLE are shown solid (magenta) and dashed (green), and unstable eigenvector in yellow. Red dots indicate initial points from where the stable manifold (blue) was continued.
 {\bf B:} Superposition of the two continued separatrix  surfaces.
 The continuation in A and B is performed as described in the Appendix.
 {\bf C:} Trajectories along the separatrix surfaces originating from SD and DS saddles points on the $S_1$ and $S_2$ manifolds, respectively.
 {\bf D:} Trajectories (dotted) along the separatrix surfaces, continued from saddle chimeras. Parameters are $\beta=0.025$ and $A=0.1$ ({\bf A-C}) or $A=0.2$ ({\bf D}).
 }
\end{figure*}

The stable manifold of the saddle chimera defines the boundary of the basin of attraction of the corresponding stable chimera.  By approximating this manifold, we can visualize which regions of the state space will evolve toward this chimera state, as shown in Figure~\ref{SI_Separatrices1} and Figure~\ref{SI_Separatrices2}.  The manifold can be approximated as follows:

{\bf Step 1:} Compute the two stable eigenvectors of the saddle chimera to obtain a local approximation to the stable manifold near the saddle chimera.  (There are two stable eigenvectors $\mathbf{v}_1$ and $\mathbf{v}_2$ and one unstable eigenvector $\mathbf{v}_3$ for the saddle chimera located at $\mathbf{p}$. The stable eigenvectors define a plane tangent to the stable manifold.) 

{\bf Step 2:} Obtain a family of starting points $\mathbf{x}_0(\theta)$ for continuation by making small perturbations off of the saddle chimera in every direction within the stable manifold. (Define a vector of angles $\theta$ and a magnitude $\epsilon$. The family of starting points are defined by $\mathbf{x}_0(\theta)=\mathbf{p}+\epsilon\left[\cos(\theta)\mathbf{v}_1+\sin(\theta)\mathbf{v}_2\right]$.) 

In Figure~\ref{SI_Separatrices1}, we used 23 angles $\theta$ between 0 and $\pi$ with the vectors $\mathbf{v}_1$ and $\mathbf{v}_2$ chosen so that all of these perturbations led to relevant parameter values and a perturbation magnitude of $\epsilon=10^{-6}$. In Figure~\ref{SI_Separatrices2}, 94 trajectories were used.

{\bf Step 3:} Integrate backward in time from each point until the trajectories reach $\mathbf{x}_1$ with a predetermined distance  $||\mathbf{x}_1-\mathbf{x}_0||$ from the start point, and plot the surface containing these trajectories. (We used \verb|ode45| to integrate the equations and then interpolated to determine when the trajectories had reached the desired length.) 

In Figure~\ref{SI_Separatrices1}, the predetermined distance was set at 0.01 to obtain a high resolution near the manifolds.  In Figure~\ref{SI_Separatrices2}, the predetermined distance was set between 1 and 20 (and no additional refinement was performed) in order to reduce data points for quick rendering, and in particular to enhance the visibility of the separatrices while reducing the total number of points displayed. This way the point families are equidistant in space, rendering an accurate picture of the separatrix surface in all regions.

{\bf Step 4:} The endpoints of the trajectories define a curve. Use evenly spaced points along the curve as new starting points and return to step 3 until enough of the stable manifold has been computed. 

In Figure~\ref{SI_Separatrices1}, we used a spacing of 0.01 near the saddle chimera and 0.05 once the trajectories had reached a distance of 0.2 from the saddle chimera.

While this method yields satisfactory results for the problem at hand, we mention that more advanced and accurate continuation methods are available for the computation of the manifold, for an overview of such methods see~\cite{Krauskopf2005}.

\section{Alternative coordinate representation}
In the main text, we chose to use the parametrization with $\rho_1,\rho_2$ and $\psi$, because this allows for visualization in a familiar cylindrical coordinate system, and because these coordinates have natural interpretations in terms of the distributions of phases in the finite oscillator system: $\rho_1$ and $\rho_2$ indicate the degree of synchrony in each population, and $\psi$ defines the mean phase difference between the populations. However, an alternative coordinate representation is possible that better reflects the symmetries inherent to the system, as is discussed here.

\begin{figure}[htp]
  \centering 
  \includegraphics[width=0.9\textwidth]{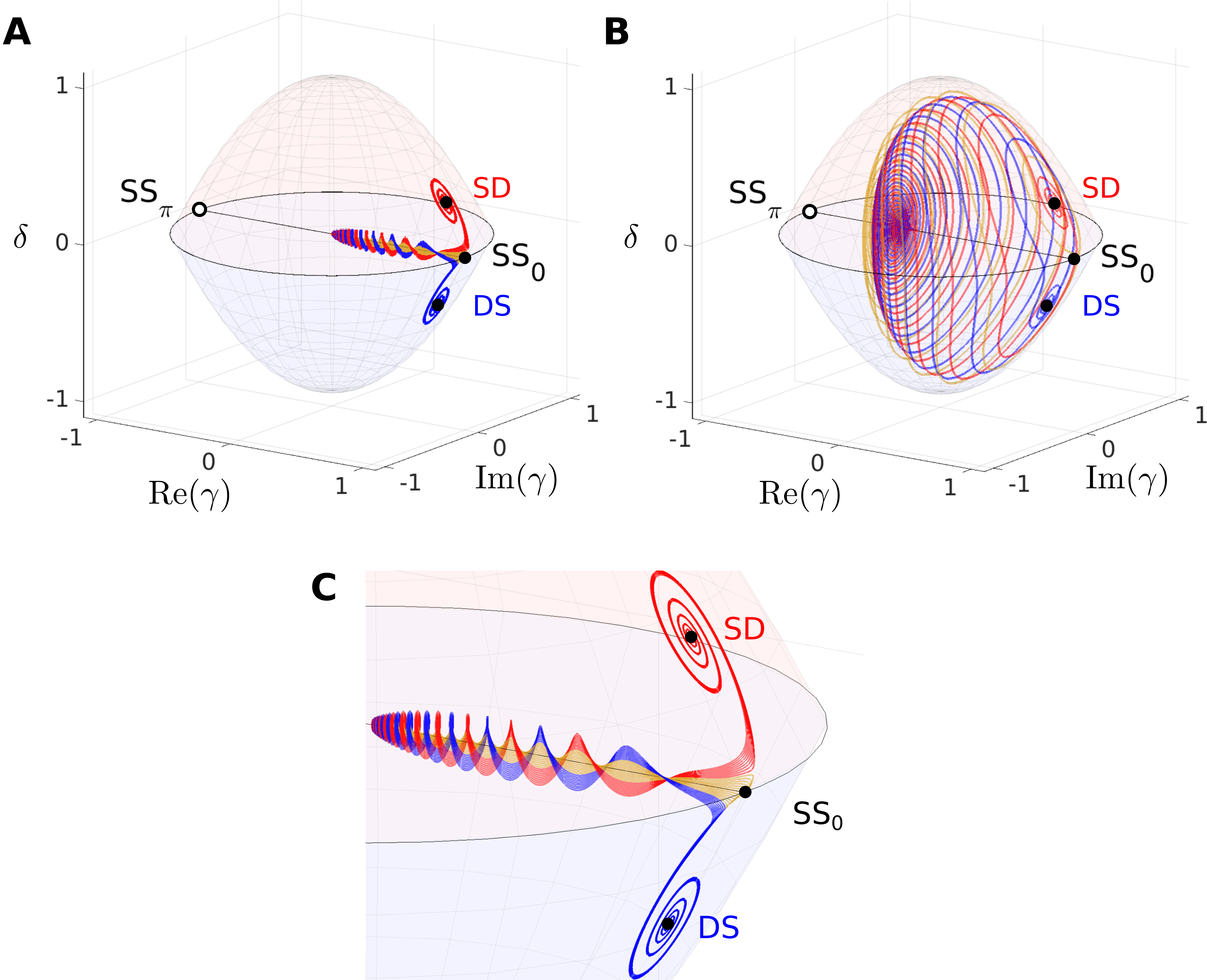}
  \caption{\label{fig:cd_coordinates}
  Trajectories with initial conditions near \Rzero\, ({\bf A}, close up in {\bf C}) and near \Rpi\, manifold ({\bf B}), leading to the states SD (red), DS (blue) and \SSzero\, (yellow). Parameters are $A=0.1$ and $\beta=0.025$. The invariant surface manifolds $S_1$ and $S_2$ are colored red and blue, respectively. 
  }
\end{figure}

The equations describing the thermodynamic limit~\eqref{eq:rho1rho2psi_eqnA}-\eqref{eq:rho1rho2psi_eqnC}, before being transformed into polar coordinates, 
can be rewritten in terms of two complex amplitudes, $\bar{z}_k=\rho_ke^{i\phi_k}, k=1,2$, taking the form
\begin{eqnarray}
 \frac{\partial \bar{z}_1}{\partial t}&=&
 \frac{\mu}{2}\left[e^{i\alpha}\bar{z}_1 -e^{-i\alpha} z_1\bar{z}_1^2\right]
+\frac{\nu}{2}\left[e^{i\alpha}\bar{z}_2 -e^{-i\alpha} z_2\bar{z}_1^2\right]
 \
\end{eqnarray}
and the corresponding equation for $z_2$ with interchanged indices.
This system exhibits a rotational symmetry according to
\begin{eqnarray*}
(z_1,z_2)&\rightarrow& (z_1 e^{i \phi}, z_2 e^{i \phi}).\
\end{eqnarray*}
This symmetry motivates a reduced coordinate system 
\begin{eqnarray}
\gamma &= z_1 \bar{z_2} \in \mathbb{C}\\
\delta &= |z_1|^2-|z_2|^2 \in \mathbb{R}\
 \end{eqnarray}
with $0\leq |\gamma|\leq 1$, and $\delta\in [-1,1]$,
from which we recover the original variables with the intuitive meaning
\begin{eqnarray*}
\rho_1^2&=&|z_1|^2 = \half (\sqrt{\delta^2+4|\gamma|^2}+\delta)\\
\rho_2^2&=&|z_2|^2 = \half (\sqrt{\delta^2+4|\gamma|^2}-\delta)\\
\psi &=& \arg{(z_1-z_2)} = \arg{c},  \
\end{eqnarray*}
provided that $|z_1|$ and $|z_2|$ are non-zero. 
This system is singular only at $z_1=z_2=0$ and its geometry can be presented so that  the symmetry of exchanging $z_1$ and $z_2$ is maintained, i.e., the reflection symmetry
\begin{eqnarray*}
 \delta &\rightarrow& -\delta\\
 \gamma &\rightarrow& \bar{\gamma}.\
\end{eqnarray*}
For this parameterization, the fully synchronized states SS$_0$ and SS$_\pi$ are located at $(\gamma,\delta)=(1,0)$ and $(\gamma,\delta)=(-1,0)$, respectively. 
The invariant rays \Rzero\, and \Rpi\, are located on the same straight line given by $\delta=0$ with $\Im{\gamma}=0$. The invariant manifolds $S_1$ and $S_2$ are the two paraboloids defined via $\pm \delta=1-|\gamma|^2$. States of interest are then in the region enclosed by these two paraboloids. Sample trajectories are shown in Fig.~\ref{fig:cd_coordinates}.

\section*{References}

\bibliographystyle{unsrt}
\def\urlprefix{}
\def\url#1{}

\end{document}